\documentclass[graphicx,amssymb, amsmath,preprint, 12pt]{revtex4-1}
\pdfoutput=1 
\usepackage{graphicx}
\usepackage{upgreek} 
\usepackage{amssymb}  
\usepackage{textpos}  
\usepackage{hyperref}
\hypersetup{colorlinks=true, citecolor=blue, urlcolor=blue, linkcolor=blue}      
\begin{document} 
\begin{textblock*}{\textwidth}(0cm,0cm)
{This article may be downloaded for personal use only. Any other use requires prior permission of AIP Publishing. {\it This is the accepted author version of the article} that appeared in Journal of Applied Physics vol. 114, 033504 (2013) [$\copyright$ 2013 AIP Publishing LLC]. The version of record is available from the publisher's website from the link: \url{https://aip.scitation.org/doi/pdf/10.1063/1.4811521}}  
\end{textblock*}  
\vspace{4cm}  
\title{Compact photonic-crystal superabsorbers from strongly absorbing media} 
\author{G. C. R. Devarapu}
\affiliation{School of Physics, College of Engineering, Mathematics and Physical Sciences (CEMPS), \\University of Exeter, Exeter, EX4 4QL, United Kingdom\\}
\author{S. Foteinopoulou} 
\email{S.Foteinopoulou@exeter.ac.uk}  
\affiliation{School of Physics, College of Engineering, Mathematics and Physical Sciences (CEMPS),
\\University of Exeter, Exeter, EX4 4QL, United Kingdom\\} 
  
\begin{abstract}     
We present a route to near-perfect absorption in compact photonic-crystal (PC) structures constructed from strongly absorbing media that are typically highly reflective in bulk form. Our analysis suggests that the key underlying mechanism in such PC superabsorbers is the existence of a PC-band-edge reflectionless condition. Although the latter is by default uncharacteristic in photonic crystals, we propose here a clear recipe on how such condition can be met by tuning the structural characteristics of one-dimensional lossy PC structures. Based on this recipe we constructed a realizable three-layer SiC-${\textrm{BaF}_{2}}$-SiC PC operating within the Reststrahlen band of SiC. We demonstrate near-perfect absorption in this prototype of total thickness smaller than $\lambda/3$, where more than $90\%$ of the impinging light is absorbed by the top deep-subwavelength layer of thickness $\sim \lambda/1100$.  We believe our study will inspire new photonic-crystal-based designs for extreme absorption harnessing across the electromagnetic spectrum. 
\end{abstract}
\maketitle 
\section{Introduction} 
\par
Absorbers are crucial components in photovoltaic and bolometric light detectors \cite{photo1, handofoptics, semiconductor1}, and thus invaluable for a wide range of applications such as  energy conversion systems \cite{atwater_review, bronger}, IR imaging devices \cite{irimag1, irimag2} for early-stage cancer diagnosis \cite{cancer1, cancer2, cancer3} as well as bio-sensing \cite{bio, mason1,mason2}. This vast applications potential has spurred intensive research efforts for new efficient absorber designs across the EM spectrum. Traditional architectures may involve a top anti-reflection coating \cite{semiconductor1} to enhance the in-coupling of light and a back-reflector that facilitates a second  light pass \cite{handofoptics}. Many current works go beyond the latter approach with focused efforts around absorption optimization by nano/micro-structuring the absorber and/or its environment \cite{gangchen, fan, fan2, postigo}, including structures aiming for plasmon-mediated near-field enhancement in the vicinity of the absorber \cite{atwater_review, bronger, aydin, mason1, mason2, cancer1}.  
\par 
Photonic-crystals have been researched for absorption control both as back-reflector components\cite{biswas, joann, ralf} and directly as the absorptive medium \cite{lin, veronis, povinelli, seassal, opex}. The latter cases seem promising schemes for one-step absorption platforms where the lossy photonic crystal could facilitate the in-coupling of all impinging light and at the same time would mold the coupled mode in a fashion that enables all light to get absorbed. This one-step process is highly attractive, but also particular challenging, even more so for thin sub-wavelength structures. Compact PC-based superabsorbers should demonstrate a strong power-loss rate; so they should be constructed from strongly absorbing media. Yet strong absorbers, i.e. media with a large extinction coefficient $\kappa$, are typically highly reflective as bulk materials. In Ref. \onlinecite{opex}, Devarapu and Foteinopoulou derived theoretically a condition for zero reflection at the interface of a lossy one-dimensional (1D) PC. Relying on this condition they demonstrated a SiC-air PC paradigm which is reflectionless even within the Reststrahlen band of SiC. This feature had been subsequently utilized to achieve a near-perfect absorption with a thick structure of about $\sim 20 \lambda$.
\par 
In this paper, we investigate control of the spectral occurrence of the aforementioned reflectionless condition with the structural characteristics of 1D PCs. Relying on the insight gained from the analysis of the latter design principle we propose compact realizable PC structures exhibiting dramatic absorption enhancement. In particular, in Sec. II we explore how the PC's structural features should be tuned for the spectral control of the reflectionless condition. In Sec. III we analyze the key importance of the spectral position of the reflectionless condition with respect to the PC band-edge on extra-ordinary absorption control. By applying the insight gained by this analysis, we explore extreme absorption harnessing with a compact three-layer PC in Sec. IV. Finally, we present practically realizable compact designs in Sec. V and discuss our conclusions in Sec. VI. 
\par
\section{Photonic Band-edge and reflectivity}
\par
In Ref. \onlinecite{lin} S. Y. Lin et al. presented a metallic photonic crystal absorber, where the observed optimal -close to $50 \%$- absorption was attributed to the low energy velocity, $\textrm v_{\textrm e}$ at the PC band-edge, providing longer light-matter interaction times. However, very recently the theoretical analysis \cite{opex}  of Devarapu and Foteinopoulou uncovered a complex relation between energy velocity and reflection. They showed that actually optimal absorption occurs when the PC is reflectionless, which is atypical for band-edge frequencies. In this section, we explore whether it is possible by tuning the PC's characteristics to push such reflectionless condition very close to the band-edge.
\par
For this purpose, and for completeness, we briefly recap the reflectionless condition  that was derived in Ref. \onlinecite{opex}. For a one-dimensional lossy PC to be reflectionless, the energy velocity at its interface should be equal to:
\begin{equation}
{\textrm v}_{\textrm{e0}}= \frac{2c}{\varepsilon^{'}+\frac{ 2 \omega \varepsilon^{''}}{\Gamma}+1},
\end{equation} 
with $c$ being the speed of light. The quantity $\varepsilon^{'}+2\omega\varepsilon^{''}/\Gamma$ appearing in the denominator of Eq. (1) is characteristic of a Lorentzian absorber \cite{loudon} of dispersive permittivity $\varepsilon(\omega)$, with $\omega$ and $\Gamma$ being the impinging electromagnetic (EM) wave's and material's damping frequency respectively. This factor is correlated with the lossy medium's stored electric energy density \cite{ruppin}. Note that throughout this paper the prime and double-prime will designate the real and imaginary part of the permittivity respectively. Eq. (1) suggests that the optimal energy velocity value at the interface that is required for the PC structure to be reflectionless is dictated only by the lossy material that is placed at the interface. However,  tailoring the energy velocity at the interface of the PC does depend highly on all its constituents and its structural particulars. 
\par
For the PC to be reflectionless, Eq. (1) is a necessary but not a sufficient condition. In addition, the impinging light must be slowing down as it enters the PC with a negative spatial energy velocity gradient that is given by:\\
\begin{equation}
\left({\frac{d {\textrm v}_{\textrm e}}{dx}}\right)_{\textrm{0}}=  \frac{\varepsilon^{''} \omega}{\textrm c} {\textrm v}_{\textrm{e0}} \left[-1+ \frac{{\textrm v}_{\textrm{e0}}}{c}\right], 
\end{equation}
where ${\textrm v}_{\textrm{e0}}$ is the required optimal interface-energy-velocity value given by Eq. (1). 
\par
\par 
\begin{figure}[htbp]
\begin{center}  
\includegraphics[angle=0, width=9.0cm]{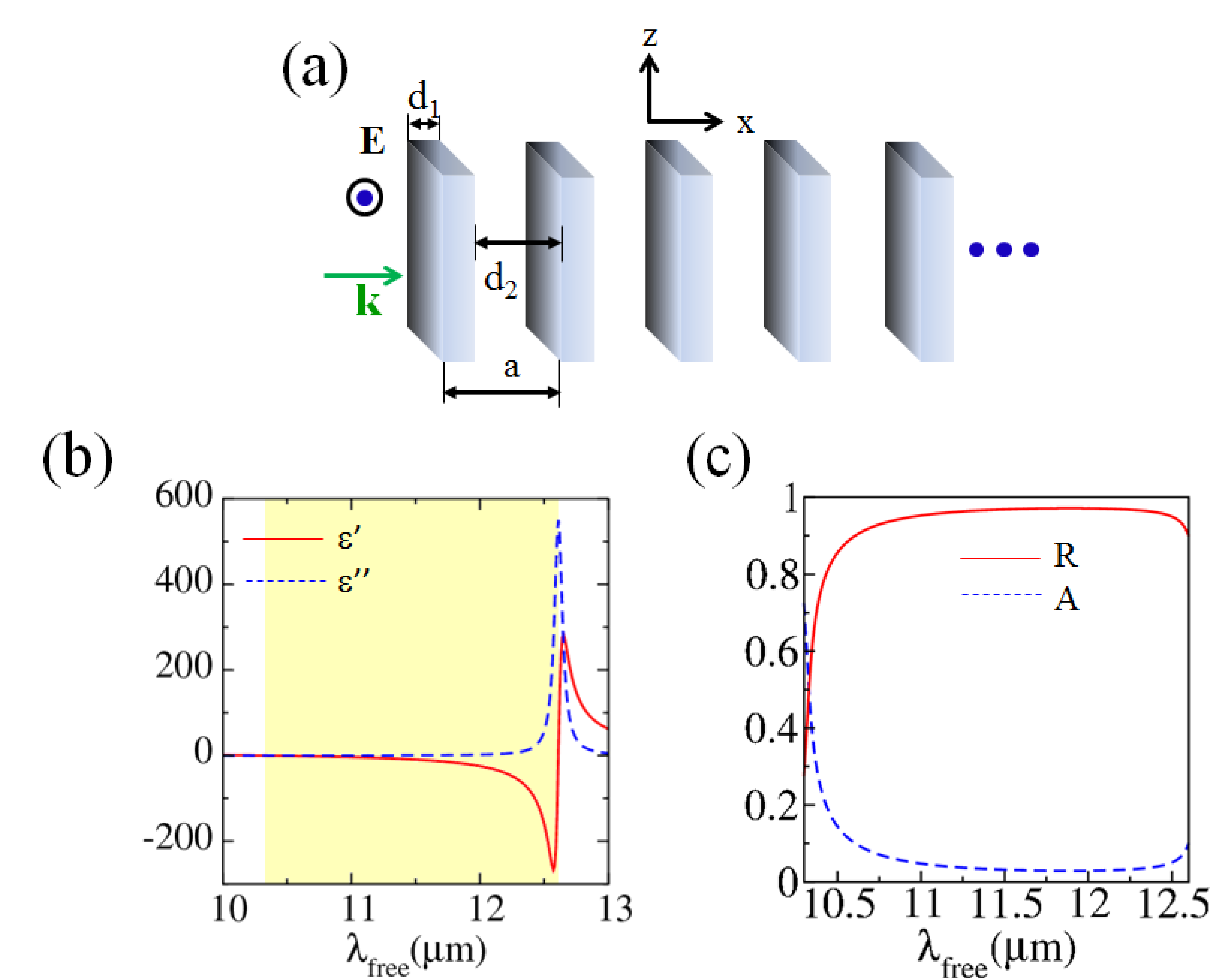} 
\vspace{1mm}
\caption{(Color online) (a) Schematics of the SiC-air 1D-PC with the geometric parameters indicated. (b) Spectral response of the real (solid) and imaginary (dashed) parts of  the SiC permittivity model of Eq. (3). (c) Absorption (solid line) and reflection (dashed line) for a thick bulk SiC block.}
\end{center}    
\end{figure}
We proceed by studying a SiC-air 1D PC system as in Ref [\onlinecite{opex}], depicted in the schematics of Fig. 1(a) with all relevant geometric features designated in the figure- with ``1'' and ``2'' labeling the properties of the SiC and air layer respectively. The corresponding permittivities are $\varepsilon_1=\varepsilon(\omega)$ and $\varepsilon_2=1$, with the dispersive dielectric  $\varepsilon(\omega)$ given by:
\begin{equation}
\varepsilon(\omega)=\varepsilon_{\infty}\hspace{0.2mm} \left(1+\frac{\omega_L^2-\omega_T^2}{\omega_T^2-\omega^2-i\omega \Gamma}\right),
\end{equation}
where $\varepsilon_{\infty}$=6.7, $\omega_T=$2$\pi \times$23.79 THz, $\omega_L=$2$\pi \times$29.07 THz and $\Gamma=$2$\pi \times$0.1428 THz in accordance with Ref. \onlinecite{sic_parameters}. Eq. (3) with these parameters give a permittivity model close to the experimental optical Palik data \cite{palik} for SiC that is appropriate for EM waves varying as $e^{-i\omega t}$ with time, t\cite{voula1, voula2}. The SiC permittivity is shown in Fig. 1(b), where we have highlighted the Reststrahlen band where our subsequent investigation focuses. In the Reststrahlen regime, light does not couple efficiently inside the bulk material. It gets reflected with very little light getting absorbed as one can see in Fig. 1(c). 
\par  
For the SiC-air 1D PC there are two main structural characteristics that can be tuned to control its behavior. One is its inter-layer separation, $\textrm a$, known as lattice constant, and the other the SiC filling ratio, $f$, given by ${\textrm d_1}/{\textrm a}$. We focus in the following only on a low filling ratio PC, with $f=0.05$. This is because we found that as the PC's filling ratio increases it start resembling the behavior of bulk SiC \cite{opex}. So the primary ``tuning knob'' to control the PC's properties will be the lattice constant, $\textrm a$. We alert the reader that the familiar scalability law applicable to dielectric PCs \cite{joannbook} does not extend to our PC system under investigation, because it is made of dispersive constituents. In other words, changes in the PC's lattice constant can in principle effect very different PC behaviors in contrast with a mere frequency shift of the same behavior which one expects for dielectric PCs.
\par
Accordingly, we investigate in the following the interface energy velocity $\textrm v_{\textrm{e, int}}$ with changing lattice constant $\textrm a$ for a semi-infinite PC \cite{semi-infinite}. It can be shown that it can be evaluated by the following expression \cite{ruppin}:
\begin{equation} 
{\textrm v}_{\textrm{e,int}}=
\frac{\frac{1}{2}  \textrm {Re}[E_y(0) H_z^*(0)] }{\frac{1}{4}  [\varepsilon_0 \left(\varepsilon^{'}+\frac{ 2 \omega \varepsilon^{''}}{\Gamma}\right) |E_y(0)|^2  \hspace{0.1mm}+ \hspace{0.1mm} \upmu_0 |H_z(0)|^2] },  
\end{equation}
where $\varepsilon_0$ and $\mu_0$ are the vacuum permittivity and permeability respectively. $E_y$ and $H_z$ represent the relevant $y-$ and $z-$ components of the electric and magnetic field [see Fig. 1(a)].  The fields are evaluated at the interface, i.e at $x=0$, with the the Transfer Matrix Method (TMM) \cite{tmm1,tmm2,tmm3}.
\par 
\par
\begin{figure}[!htbp]
\begin{center}  
\includegraphics[angle=0, width=7.5cm]{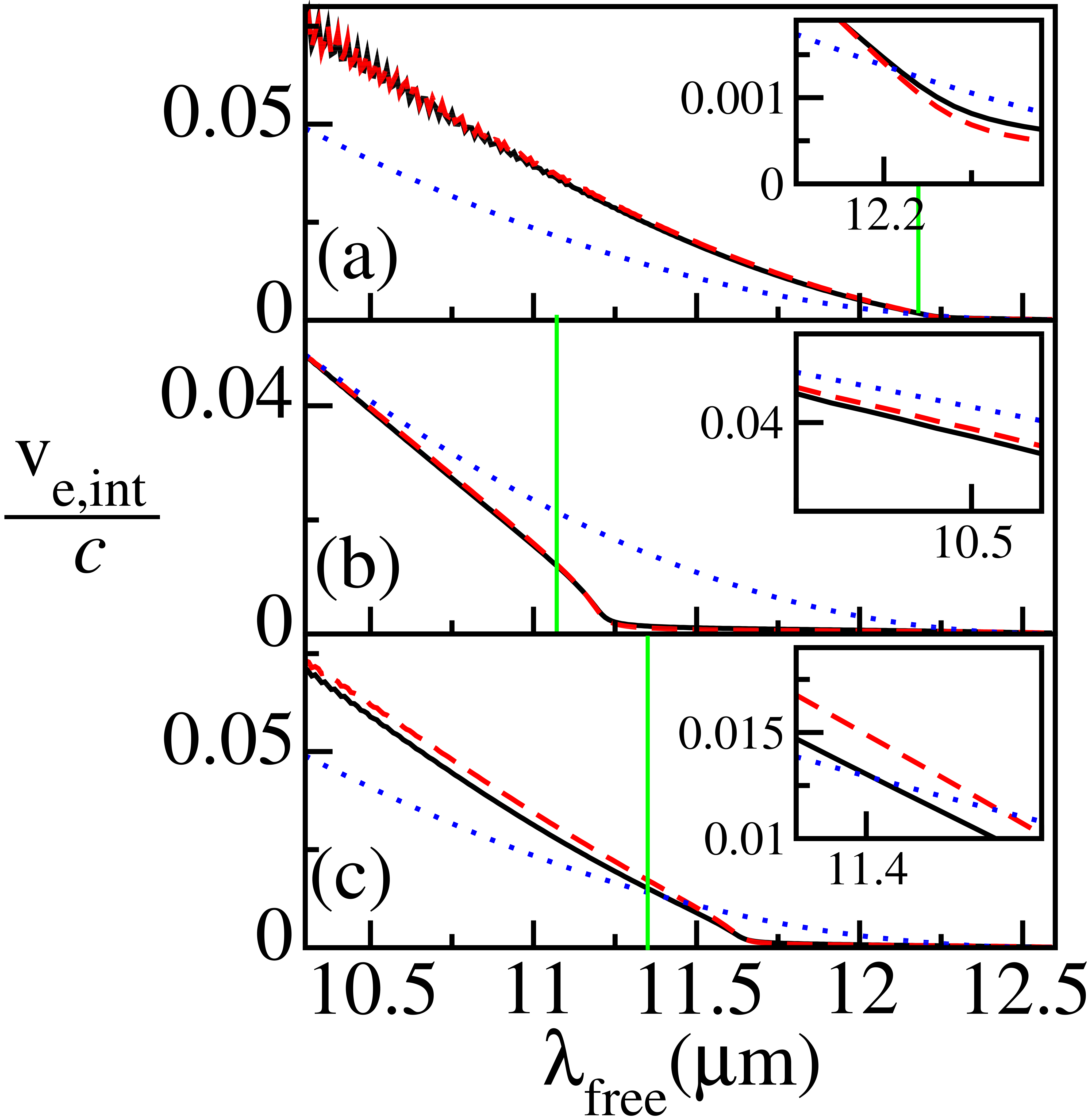}
\caption
{(Color online) Spectral response of the energy velocity at the interface ${\textrm v}_{\textrm{e,int}}$ of a semi-infinite SiC-air PCs structure is shown as solid lines. The dashed lines depict the corresponding values for the same PCs but with $50\%$ of their entry face being cut-off. The results in (a), (b) and (c) represent the PC cases with a lattice constant of a equal to 5 $\upmu {\textrm m}$, 8 $\upmu {\textrm m}$ and 10 $\upmu {\textrm m}$, respectively. In all cases, the interface-energy velocity value of the reflectionless condition, ${\textrm v}_{\textrm{e,0}}$ of Eq. (1), is depicted with dotted lines. Note, all energy velocity values are expressed in terms of the speed of light $c$. The vertical solid lines represent the spectral position of the absorption peaks that we will observe in Fig. 5.}
\end{center}  
\end{figure} 
\par
\par
Three characteristic cases for the spectral response of the energy velocity at the PC interface, $\textrm v_{\textrm{e,int}}$, are shown as solid lines in Figs. 2 (a), 2(b) and 2(c), that correspond to PCs with lattice constants a=5 $\upmu{\textrm m}$, 8 $\upmu{\textrm m}$  and 10 $\upmu{\textrm m}$ respectively. Note, $\textrm v_{\textrm{e,int}}$ is expressed as a fraction of the speed of light, $c$ and is plotted versus the free space wavelength $\lambda_{\textrm{free}}$ of the incoming EM wave. We can identify the PC band-edge in each case of Fig. 2 with the region where the energy velocity drops to nearly zero, but we note that this is not sharply defined for lossy media. We also show the reflectionless condition by plotting the required interface-energy-velocity optimum, ${\textrm v}_{\textrm{e,0}}$,  with the dotted lines. The intersection between the ${\textrm v}_{\textrm{e,int}}$ and the ${\textrm v}_{\textrm e,0}$ curves signifies the free-space wavelength where the PC can be reflectionless. We highlight more clearly this region in the insets of each figure. 
\par 
We observe three distinct behaviors with respect to the relation between the ${\textrm v}_{\textrm{e,int}}$ and ${\textrm v}_{\textrm e,0}$ curves. The eight-micron structure case, seems to be completely different from the case of the five- and ten-micron structures. In particular, we see that the two energy velocity curves scrape close together without intersecting, at an extended wavelength region around 10.5 $\upmu$m and far from the band-edge. Based on this observation, we predict that the eight-micron design would not be the most suitable for an enhanced absorption performance, which we will confirm in Sec. III. The behavior of the five-micron and ten-micron designs looks similar where we see that the $\textrm v_{\textrm{e,int}}$ and $\textrm v_{\textrm{e,0}}$ curves intersect, which means both can potentially operate as reflectionless PCs at the intersection wavelength. Yet, there is a small, but very important difference. For the five-micron PC design the intersection between  $\textrm v_{\textrm{e,int}}$ and $\textrm v_{\textrm{e,0}}$ curves occurs at the close vicinity of the band-edge. On the other hand,  this intersection occurs somewhat further from the band-edge for the ten-micron PC design, whose behavior is similar to the case studied in Ref. \onlinecite{opex}. We will see in the following section, that even this seemingly small spectral difference in the intersection occurrence with respect to the band-edge will have an enormous impact on the respective PCs performances as absorbers. 
\par  
We would like to remind the reader, that the existence of an intersection between  ${\textrm v}_{\textrm{e,int}}$ and ${\textrm v}_{\textrm e,0}$ is only a necessary condition. Light should also slow down as it enters the PC structure with an appropriate spatial gradient given by Eq. (2). Altering only the entry-layer thickness, without modifying the remaining PC characteristics,  tunes the interface energy velocity gradient \cite{opex} without changing much the value of  ${\textrm v}_{\textrm{e,int}}$. We can see that by observing the dashed line curves in Fig. 2, which correspond to the same semi-infinite PCs, but with $50\%$ of their entry SiC layer being cut-off. 
\par
\begin{figure}[htbp]
\begin{center}  
\includegraphics[angle=0, width=6.0cm]{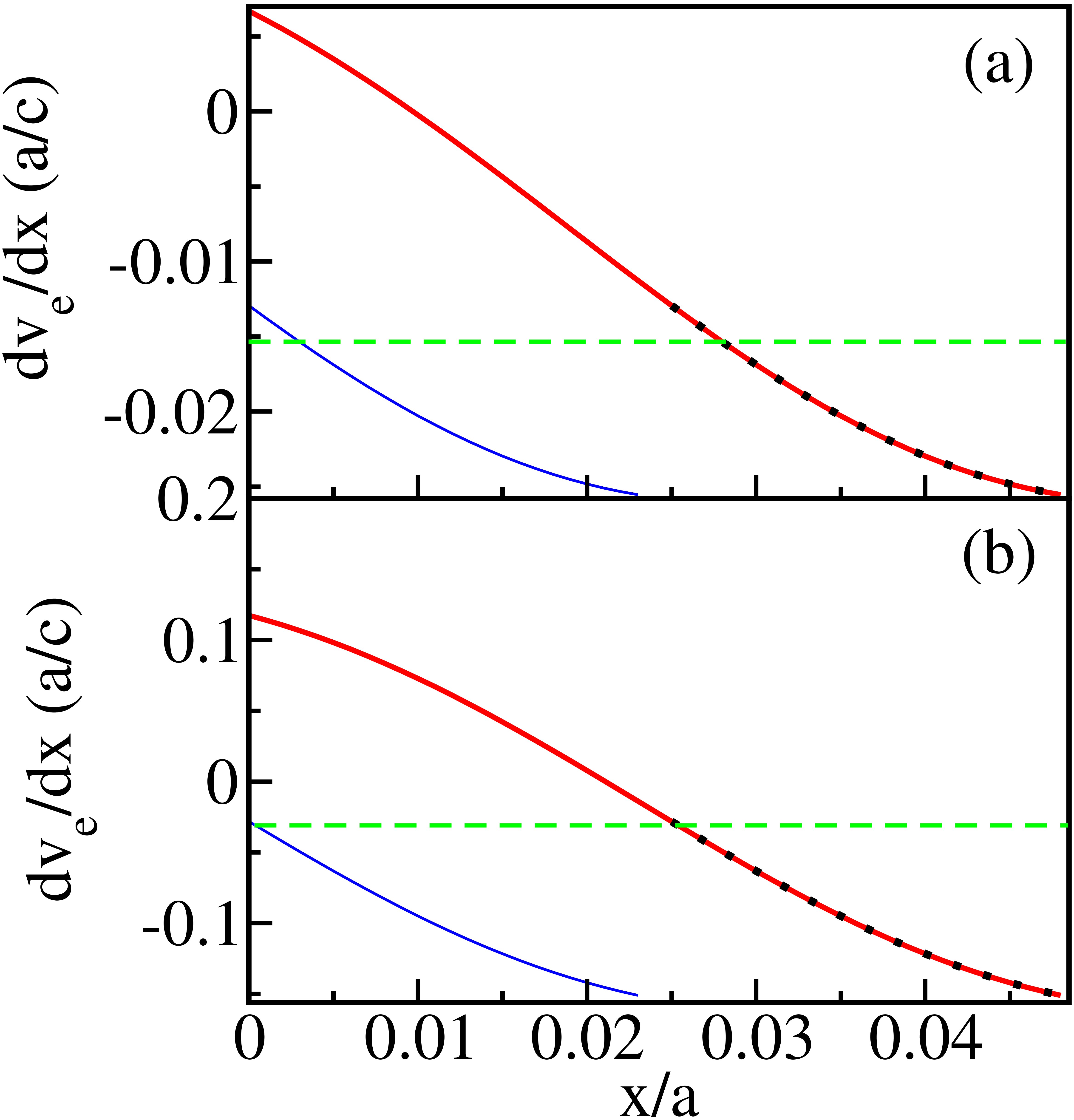}
\caption
{(Color online) The energy-velocity gradient is shown for two PC systems with a lattice constant equal to 5 $\upmu {\textrm m}$ and 10 $\upmu {\textrm m}$ in panels (a) and (b) respectively. The horizontal dashed line represents the reflectionless condition value dictated by Eq. (2). Note the coordinate within the PC entry layer, $x-$, is expressed in terms of the lattice constant a,  while the energy velocity gradient is expressed in terms of $c/\textrm a$, with $c$ being the speed of light.} 
\end{center}    
\end{figure}
\par
This effect can be clearly seen in Fig. 3, where  $d{\textrm v}_{\textrm{e}}/dx$ is calculated \cite{tmm1, tmm2, tmm3} as a function of the location, $x$ within the entry SiC layer of the semi-infinite PC. This is done at the free space wavelength where the $\textrm v_{\textrm{e,int}}$ and $\textrm v_{\textrm{e,0}}$ curves intersect. This is is 12.20 $\upmu$m for the PC with a=5 $\upmu$m and 11.48 $\upmu$m for the PC with a=10 $\upmu$m. We show the results for both the cases of complete and terminated -missing  $50\%$ of the entry layer-, structures, as thick and thin solid lines respectively. Fig. 3(a) features the PC case of 5 $ \upmu {\textrm m}$ lattice constant. Conversely, Fig. 3(b) shows the corresponding result for the PC case of  10 $ \upmu {\textrm m}$ lattice constant. The dotted lines represent $d{\textrm v}_{\textrm{e}}/dx$ for the terminated PC cases, but plotted with a $x-$coordinate shift that places the terminated PC's front-layer in the center of the complete PCs  front-layer. Notice the remarkable agreement between the complete PC result and the coordinate-shifted terminated-PC result. This makes it evident why one can essentially get the desired $d{\textrm v}_{\textrm{e}}/dx$ value (seen as horizontal dashed lines), that is dictated by the reflectionless condition of Eq. (2), just by ``cutting-off'' sufficient material from the front SiC layer.
\par  
\begin{figure}[!htbp]
\begin{center}  
\includegraphics[angle=0, width=8cm]{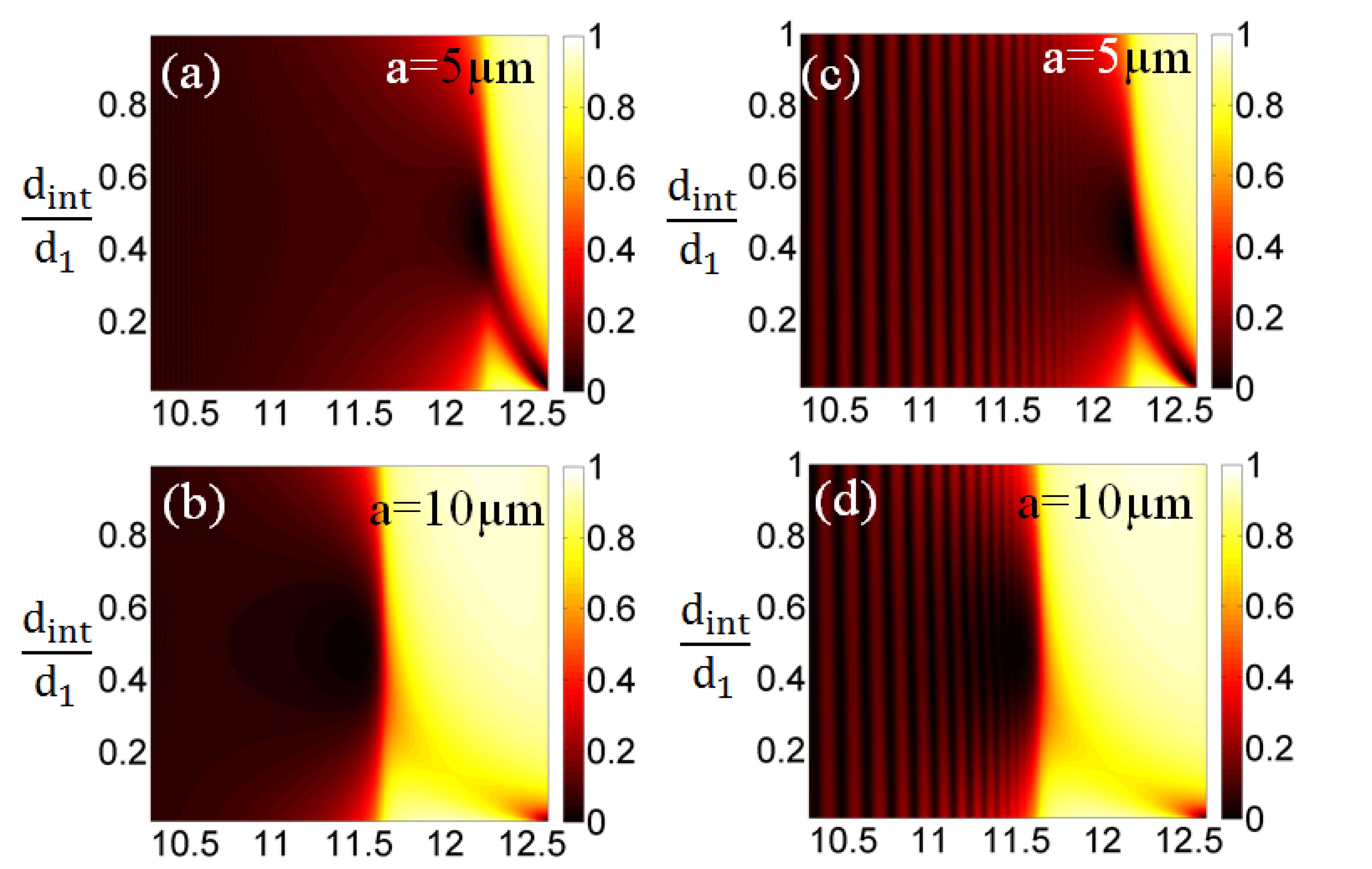}
\caption
{(Color online) Reflection (in color-map) versus termination ratio, $\textrm d_{\textrm{int}}/\textrm d_1$ and free space wavelength, $\lambda_{\textrm {free}}$ calculated from TMM. Panels (a) and (b) represent the result corresponding to the semi-infinite PCs with lattice constant a, of 5  $\upmu {\textrm m}$, and 10 $\upmu {\textrm m}$ respectively. Same is shown in (c) and (d) but for 200 $\upmu$m-thick PCs.}
\end{center}    
\end{figure}
\par
The above theoretical analysis suggests that both PCs with lattice constants of 5 and 10 $\upmu$m will be reflectionless for an impinging wave with a wavelength of 12.20 $\upmu$m and 11.48 $\upmu$m respectively, when their front-face is terminated at $44 \%$  and $50 \%$ of its complete size, respectively. To verify the predictions of our energy-velocity-based design principle we calculate with the TMM \cite{tmm1, tmm2, tmm3} the reflection for these semi-infinite PC cases within the entire Reststrahlen band and for different front-layer terminations, given by the thickness ratio of the truncated versus complete PC entry layer,  $\textrm d_{\textrm{int}}/\textrm d_1$. We show the results in Figs. 4(a) and (b) for the respective PCs of lattice constants 5 and 10 $\upmu$m. Notice, the complete agreement for near-zero reflectivity between the predictions deduced from the principles represented by Eqs. (1) and (2) and the actual TMM results. It should be noted the reflectionless PC parameters (free space wavelength and termination) are relative robust and do not change while transitioning to a finite 200 $\upmu$m-thick PC structure. For comparison, we  also show the corresponding TMM results for the 200 $\upmu$m-thick structures in Figs. 4(c) and (d).
\par
To recap, the main ``tuning-knob'' to obtain a potential near-band-edge reflectionless semi-infinite PC is the lattice constant, a. The lattice constant should be tuned, while maintaining a low PC filling ratio, so that the energy velocity at the interface equals the optimum value given by Eq. (1) at a frequency that is very close to the band-edge. Near-band-edge near-zero reflection can then be achieved, by fine-tuning the size of the entry PC layer, so that the energy velocity gradient at the interface becomes equal to the optimum value determined by Eq. (2). We will explore in the following how the aforementioned design recipe can enable extreme absorption control towards our target of compact realizable PC absorbing structures.  
\par
\section{Near-band edge near-zero reflection and absorption harnessing} 

 \par
Naturally, one would expect near-perfect absorption for the thick $200 \upmu$m-PCs of Fig. 4 at free space wavelength and termination where reflection is near-zero. To compare the two designs of 5 and 10 $\upmu$m lattice constant respectively we adopt from here-on a common $50 \%$ termination, which is near-optimum for both the cases. We calculate with TMM the absorptance, A=1-T-R, with T and R being transmission and reflection respectively and show the results in Fig. 5 as dot-dashed line for the case of a=5 $\upmu$m and solid line for the case of  a=10 $\upmu$m. 
\par
\begin{figure}[!hb]
\begin{center}  
\includegraphics[angle=0, width=7cm]{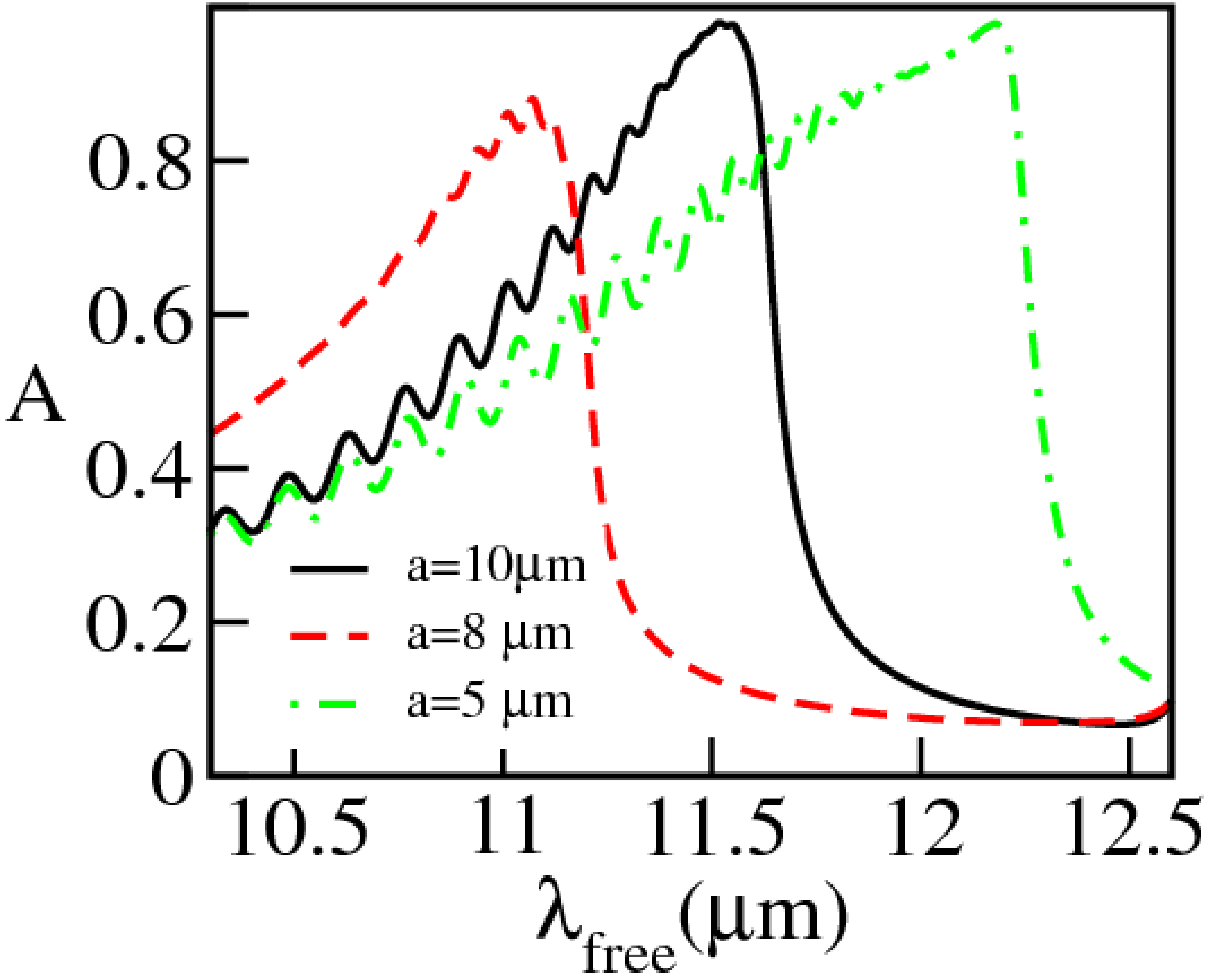}

\caption
{(Color online) Absorptance versus free space wavelength, $\lambda_{\textrm{free}}$, for three 200 $\upmu {\textrm m}$ thick SiC-air PCs of 0.05 filling ratio and $50\%$ front layer truncation. The solid, dashed and dot-dashed curves correspond to PCs with a lattice constant a equal to 10 $\upmu {\textrm m}$, 8 $ \upmu {\textrm m}$, and 5 $\upmu {\textrm m}$ respectively. The front SiC layer is terminated to half its original size.} 
\end{center}    
\end{figure} 
\par 
Indeed, we observe a near-perfect absorption for both these PC designs with spectral occurrence in excellent agreement with the intersection-wavelength prediction of Fig. 2. This excellent agreement can be easily seen by noticing the vertical solid lines in Fig. 2 designating the spectral positions of the absorption peaks observed in Fig. 5. For comparison, we also show the PC case with a=8 $\upmu$m, which did not achieve a near-perfect absorption. For the latter case, absorption peaks away from the spectral region where the ${\textrm v}_{\textrm{e,int}}$ and the ${\textrm v}_{\textrm e,0}$ curves of Fig. 2(b) are close to each other. This is because the aforementioned spectral regime, is too far from the band-edge. We checked that this remains true for any termination, and is consistent with our predictions in the previous section that this design would not be appropriate for dramatic absorption control. We therefore focus only on the 5 and 10 $\upmu$m lattice constant PC designs from thereon. 
\par
We discussed in Sec. II, that although the above mentioned PC designs show quite a similar behavior, in the sense that the ${\textrm v}_{\textrm{e,int}}$ and the ${\textrm v}_{\textrm e,0}$ curves intersect spectrally in the neighborhood of the band-edge, there is still a difference. That is such intersection being in one case much closer spectrally to the band-edge than in the other, -- roughly 0.4$\%$ and 1.4$\%$ for the respective cases of 5 and 10 $\upmu$m lattice constant. Although, on a superficial look this difference may not seem so large we will find out that it does make an enormous difference towards our goal
for compact absorbing structures.
\par
\begin{figure}[htbp]
\begin{center}  
\includegraphics[angle=0, width=8.0cm]{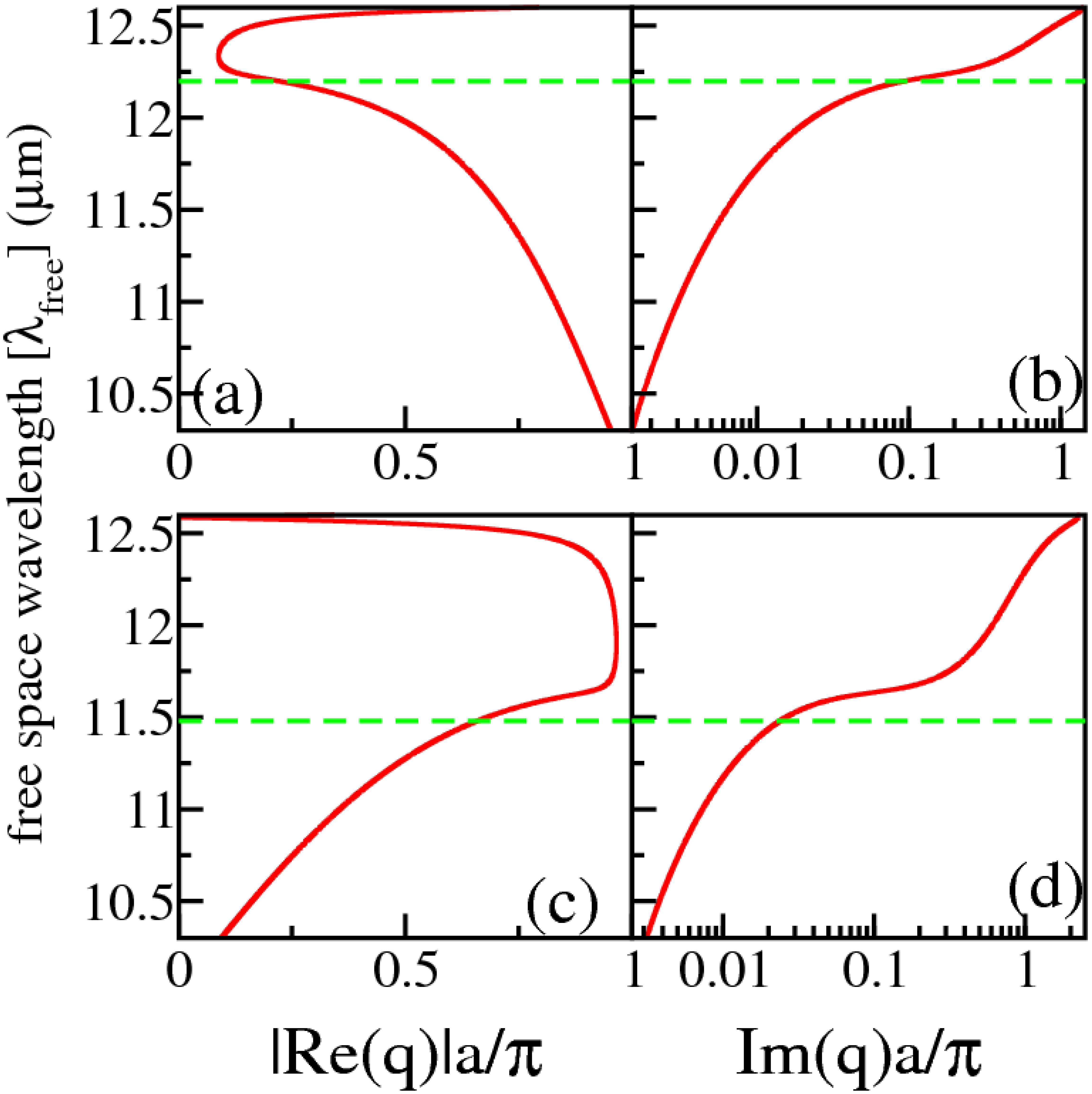}

\caption
{(Color online) Complex band structure (free space wavelength versus Bloch wavevector q) for the PC cases of lattice constant a, 5 $\upmu {\textrm m}$ [in (a) and (b)] and 10 $\upmu {\textrm m}$ [in (c) and (d)] . The respective reflectionless-condition wavelengths are indicated with horizontal dashed lines. Note, both the real and imaginary parts of the Bloch wave vector q is expressed in terms of $\pi/\textrm a$.} 
\end{center}    
\end{figure}
\par
\par
We can understand why by looking at the complex photonic band structures for the two PC cases in Fig. 6. These bandstructure calculations are performed by using the TMM and  applying the Bloch boundary conditions, with a complex  Bloch wavevector q,  \cite{tmm1,tmm2,tmm3}. The free space wavelength versus the real and imaginary parts of q is shown respectively in sub-figures (a) and (b) for the PC with a=5 $\upmu$m and in sub-figures (c) and (c) for the PC with a=10 $\upmu$m. We also indicate the ${\textrm v}_{\textrm{e,int}}$-${\textrm v}_{\textrm e,0}$ intersection wavelength of Fig.2, representing the reflectionless condition, with horizontal dashed lines. Notice, that indeed as we mentioned before the bandedges are not as sharply defined as in lossless dielectric PCs. 
\par
Fig. 6 clearly demonstrates the implications of having the reflectionless condition in the close vicinity of the band-edge. At the reflectioness condition wavelength (dashed lines) we find an $\textrm{Im(q)}=0.094 \hspace{0.01mm} \pi/\textrm a$ for the PC with lattice constant a= 5 $\upmu$m. Conversely, for the PC with lattice constant a= 10 $\upmu$m we have $\textrm{Im(q)}=0.024 \hspace{0.01mm} \pi/\textrm a$. The complex Bloch wave vector q,  implies an $e^{i{\textrm q}x}$ envelope for the electric fields spatial maps across the PC, determining a relative amplitude and phase between points with a separation that is an integer multiple of the lattice constant a\cite{longprb, complexbands}. Thus we anticipate, that the larger $\textrm{Im(q)}$ is, the quicker the electric-field amplitude decay within the PC; hence the merit of having the reflectionless condition as close to the band-edge as possible. 
\par

\par
In the following, we evaluate the above two PC candidates potential as compact absorbers. We further investigate where is the light getting absorbed while crossing the PC structure. For this purpose, we calculate the ratio of dissipated to incident power within the $\textrm j^{\textrm th}$ SiC layer, P(j), by applying Poynting's theorem \cite{jackson,aydin}:
\begin{equation}
{\textrm P}({\textrm j})= \frac{\omega \varepsilon^{''}}{{\textrm c}|E_0|^2 } \int_{x_1}^{x_2}  |E_y(x)|^2 dx,
\end{equation}
where ${\textrm x_1, x_2}$ represent the coordinate limits of the $j^{\textrm th}$ SiC layer, given by:
\begin{displaymath}
 [x_1, x_2]=\left \{\begin{array}{ll} [0, \frac{\textrm d_1}{2}] &  \textrm{ j $=$ 1 }\\
 {\textrm {[(j-1) \textrm a}}+\frac{{\textrm d_1}}{2}, {\textrm {(j-1) \textrm a}}+\frac{{\textrm d_1}}{2}]  & \textrm{ j $\not=$ 1 }.\\
\end{array}\right. 
\end{displaymath}
$E_y(x)$ represents the electric field distributions in the PC that are calculated using TMM \cite{tmm1,tmm2,tmm3} with $|E_0|$ being the incident electric field amplitude. 
\par
We briefly digress here to note that a thicker slab does not necessarily imply a higher power loss. By making the crude, -- yet reasonable for subwavelength blocks--, assumption that the electric field does not vary too much within a certain slab, and noting that $\textrm P_{{\textrm{max}}}=1$ we can obtain
from Eq. (5) an upper bound for the field enhancement $|\textrm E_{\textrm{enha}}|$. This is given by:
\begin{equation}
|\textrm E_{\textrm{enha, max}}|^2 \approx \frac{\lambda_{\textrm{free}}}{2\pi \varepsilon^{''}\textrm d_1}.
\end{equation}
Basically the above equation implies that thinner slabs may be capable of a higher field enhancement. Thus, there is no physical limitation in them yielding 
a higher power loss. In fact, we will find out exactly that in the following, namely a higher power loss in the thinner-slab five-micron PC.
\par
\begin{figure}[htbp]
\begin{center}  
\includegraphics[angle=0, width=7cm]{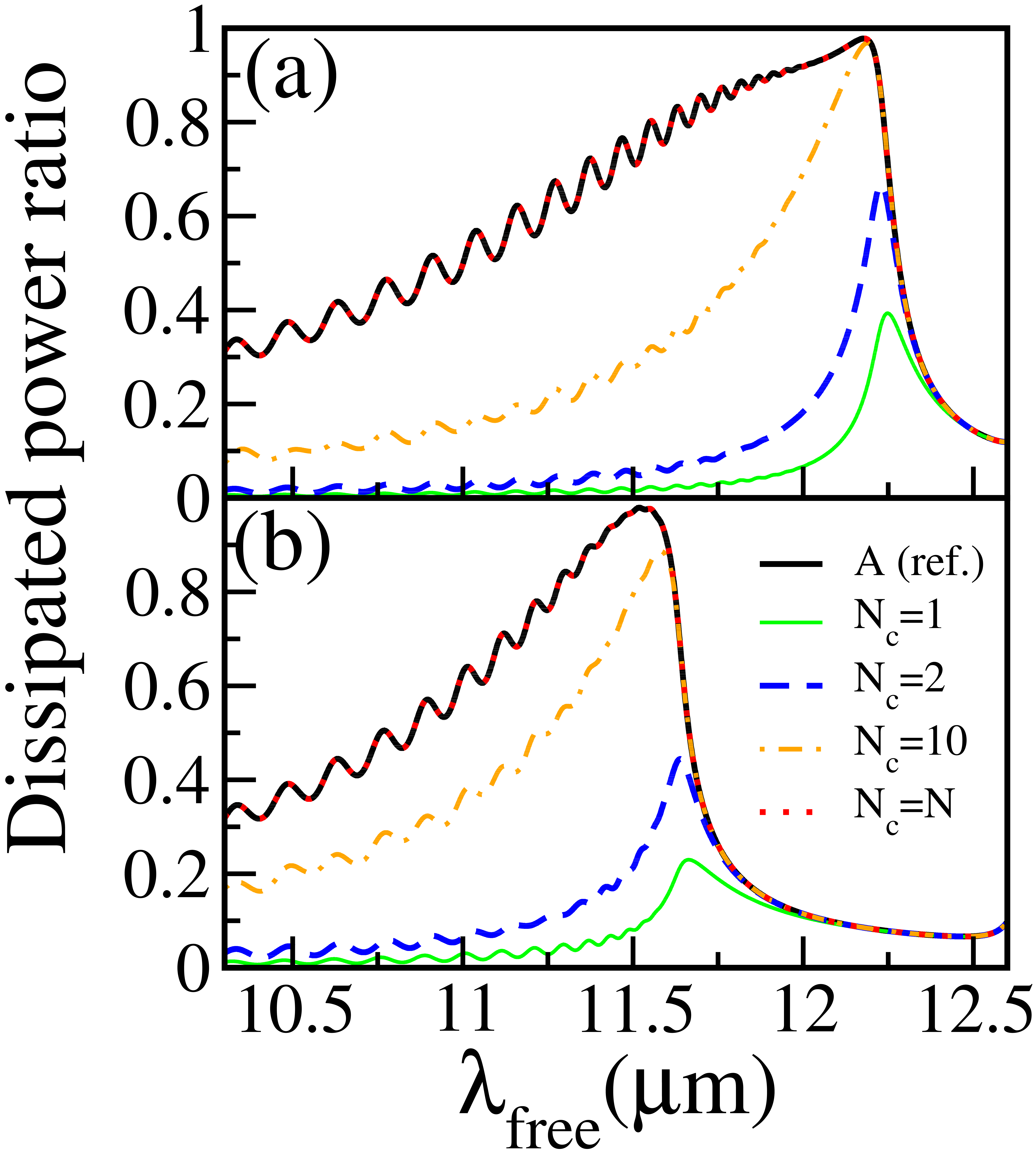}

\caption
{(Color online) Dissipated to incident power ratio versus free space wavelength, $\lambda_{\textrm{free}}$, for the 200$\upmu$m thick SiC-air PCs with $50\%$ truncated front layer, within the first $\textrm N_c$ PC unit cells. The result in (a) [(b)] corresponds to the PC case with 5 $\upmu {\textrm m}$ [10 $\upmu {\textrm m}$] lattice constant. The respective absorptance is shown for reference with the dark solid line. Note, the total number of PC unit cells, N, is 40 for the case in (a) and 20 for the case in (b).} 

\end{center}    
\end{figure}
\par
\begin{figure}[htbp]
\begin{center}  
\includegraphics[angle=0, width=6.5cm]{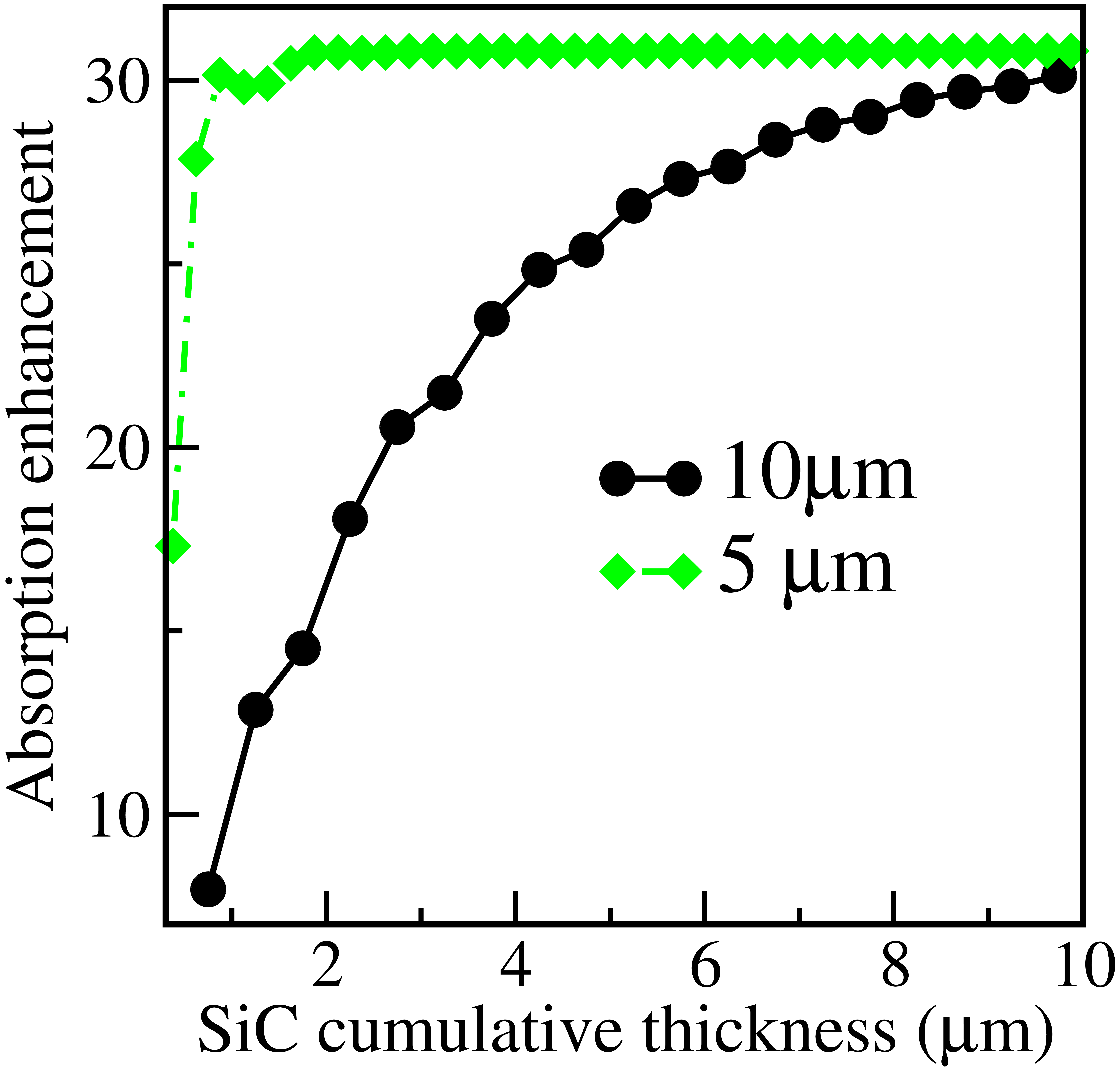}

\caption
{(Color online) Absorptance enhancement of the two terminated SiC-air PCs with  lattice constant a, 5 $\upmu {\textrm m}$ (dot-dashed line with diamonds),  and 10 $\upmu {\textrm m}$ (solid line with filled circles)  with respect to the absorption of a SiC block about a wavelength-thick is plotted against the total thickness of SiC  encountered by the EM wave as it travels through the PC.}

\end{center}    
\end{figure}
\par
We proceed in calculating what fraction of the incident power was dissipated while the EM wave has crossed through the first $\textrm N_\textrm c$ SiC layers of the PC, that we represent in the following as $\textrm P^c(\textrm N_c)$. Thus: 
\begin{equation}
\textrm P^c(\textrm N_c)= \sum_{\textrm j=1}^{\textrm {N}_\textrm c} {\textrm P} ({\textrm j}).
\end{equation}
We plot the results for $\textrm P^c(\textrm N_c)$ for the two candidate PCs; the 5 $\upmu {\textrm m}$ lattice constant case in Fig. 7(a) and the 10 $\upmu {\textrm m}$ lattice constant case in Fig. 7(b). For reference we plot also the absorptance A (dark solid line). The figure verifies, that absorptance is equal to the dissipated power through the entire PC (dotted line).  Notice, how quicker is power getting absorbed within the five-micron PC design. It is actually very impressive to observe that 60$\%$ of the impinging light at the peak wavelength is dissipated within first two SiC layers.  
\par
So, the power loss in the 5 $\upmu$m-lattice-constant PC of 200 $\upmu$m thickness is much higher than in the 10 $\upmu$m-lattice-constant one. This may imply a stronger potential of the former PC structure as a compact absorber. In order to verify this,  we test the PC's absorbing performance as it shrinks in thickness, comprising from a smaller number of unit cells.  We take the absorption given by a thick bulk slab of SiC as a measure of comparison and evaluate the absorption enhancement with respect to that given by the two PC systems under investigation. We let their respective thickness vary, while maintaining the same $50\%$ front-layer termination, and consider from compact two-unit-cell structures, to multiple-wavelength thick PCs.  We plot our results in Fig. 8, versus the cumulative SiC thickness the EM wave crosses in each case.  The dot-dashed line with diamonds designates the PC case of 5 $\upmu$m lattice constant, while the solid line with filled circles designates the PC case of 10 $\upmu$m lattice constant. Indeed, the 5 $\upmu$m-lattice constant PC yields a remarkable absorption enhancement factor over 15 even with just two unit cells, that quickly reaches saturation at a value of $\sim 30$. We will further look then into the properties of such a compact absorbing PC structure in the following section.    
\par   
\section{Compact sub-$\lambda$ PC-based absorber}
\par
Given the favorable results in Fig. 8, we take the extreme case of truncating the 5 $\upmu$m-lattice-constant PC down to only two unit cells and analyze it further. This compact design, depicted in Fig. 9(a), is essentially made of just two SiC layers that are $\textrm d_{\textrm{sp}}$ apart, with $\textrm d_{\textrm{sp}}$ being equal to $\textrm d_2$, -- the air-layer thickness of the full periodic structure depicted in Fig. 1(a). The back layer has thickness equal to $\textrm d_1$, the SiC- layer thickness of the full periodic PC, while the front layer is truncated to thickness $\textrm d_{\textrm{int}}$, that is a percentage of the back-layer thickness. 
\begin{figure}[htbp]
\begin{center}  
\includegraphics[angle=0, width=8.0cm]{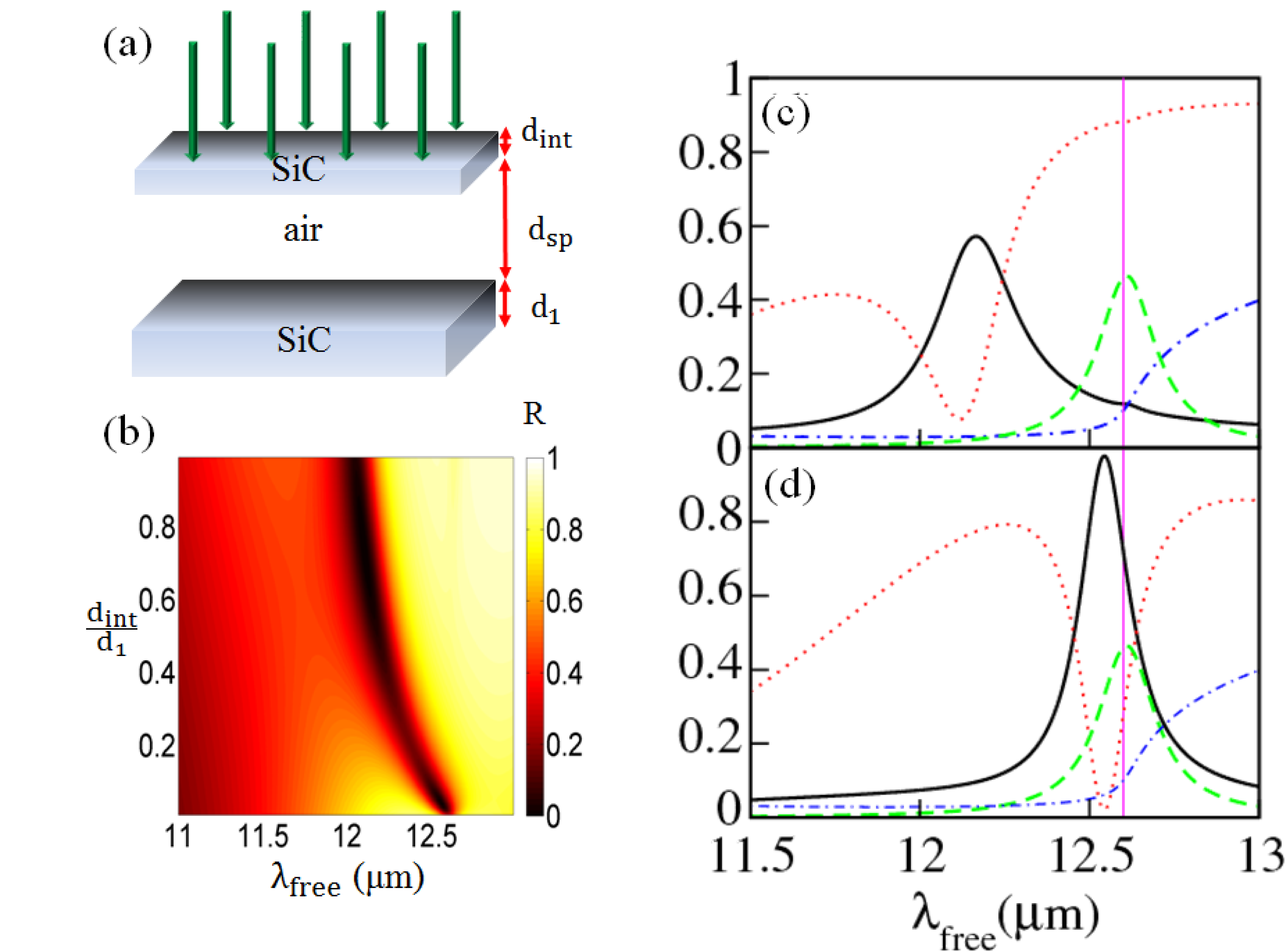}

\caption{(Color online) (a) Schematics of the compact PC-based design with all structural information indicated. (b) Reflectance (color-map) versus free space wavelength $\lambda_{{\textrm{free}}}$ and front-layer truncation ratio $\textrm d_{{\textrm{int}}}/{\textrm d_1}$. (c) Absorptance (solid lines) and reflectance (dotted lines), for the design in (a) with (c) [(d)] showing the case of $\textrm d_{\textrm{int}}/{\textrm d_1}=0.5$ [$\textrm d_{\textrm{int}}/{\textrm d_1}=0.05$]. For comparison absorptance through a single layer is also shown for bulk SiC (dot-dashed) and an ultra-thin SiC film as thick as the front layer of the structure of Fig. 9(d). The vertical line designates the SiC Reststrahlen band-edge.}

\end{center}    
\end{figure} 
\par
It is quite remarkable, how thickness-robust the PC system is. We observe in Fig. 9(b) that it retains a low reflectivity close to the original reflectionless condition of the semi-infinite system, i.e at close to 50$\%$ termination at about 12.2 $\upmu$m free space wavelength. In addition, we also observe a near-zero reflection at much smaller termination ratio ($\sim 0.05$) at about 12.50 $\upmu$m free space wavelength. If we look back at Fig. 2(a), we identify a second regime where the reflectionless condition applies. The existence of such second regime is unique to the five-micron PC design only and we did not observe it for the ten-micron PC case of Fig. 4(b). It emanates from a second intersection between the ${\textrm v}_{\textrm{e,int}}$ and ${\textrm v}_{\textrm e,0}$ curves that occurs closer to the adjacent band where $\textrm{Im(q)}$ is much larger. This second reflectionless PC regime requires much more SiC material to be ``shaved-off'' from the front layer in order to obtain the optimum energy velocity gradient of Eq. (2).
\par
\begin{figure}[htbp]
\begin{center}  
\includegraphics[angle=0, width=6.5cm]{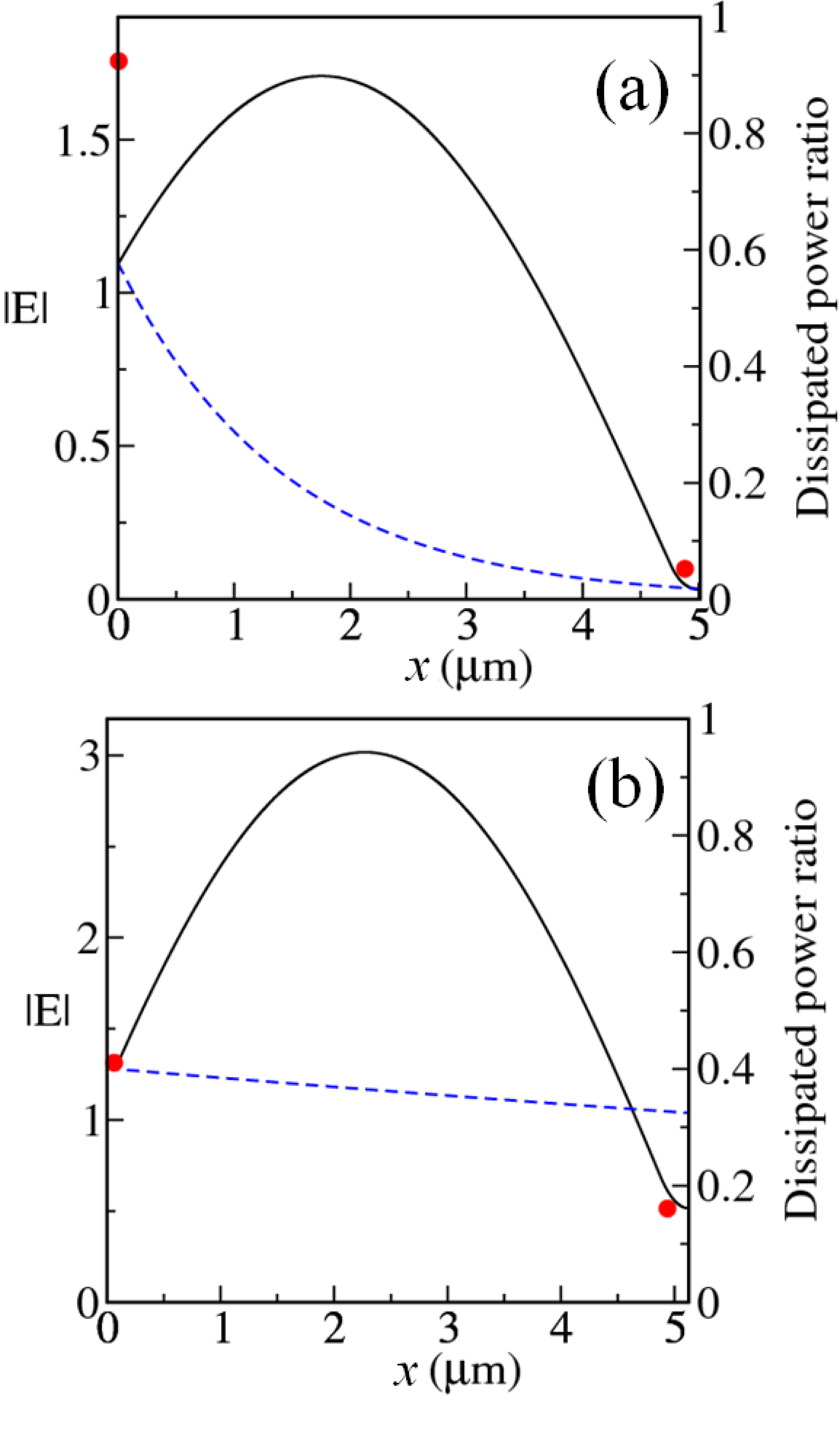}

\caption{(Color online) Electric field amplitude, $|\textrm E|$, profiles (left vertical axis) versus the coordinate $x$ within the compact superabsorber design. The depicted profiles are normalized with the incident electric field amplitude $|\textrm E_0|$. The dotted lines represent the $|\textrm E|$-decay, from the front to theback later, as predicted by the complex band structure of Fig. 6. The solid circles represent the ratio of incident power that is absorbed in each layer (see right vertical axis for values). Panel (a) and (b) represent the respective cases with front-to-back-layer truncation ratio of 0.05 and 0.5.}

\end{center}    
\end{figure} 
\par
Indeed, we can observe clearly a small reflection at about 12.2 $\upmu$m free space wavelength for the compact design of Fig. 9(a), with $\textrm d_{\textrm{int}}=0.5 \hspace{0.01mm} \textrm d_1$ seen as dotted line in Fig. 9(c). Conversely, we observe a near-zero reflection at about 12.5 $\upmu$m free space wavelength for the compact design of Fig. 9(a), with $\textrm d_{\textrm{int}}=0.05 \hspace{0.01mm} \textrm d_1$, that we show with a dotted line in Fig. 9(d). The corresponding absorptances are shown as solid lines in the same sub-figures. The absorptance peak-reflection dip offset observed in Fig. 9(c) comes from the fact that reflection is small yet non-near-zero at minimum. The peak-absorptance for the case of Fig. 9(d) correlated to the second reflectionless-PC regime is near-perfect, and much stronger in comparison to the case of Fig. 9(c). This is because the second reflectionless regime is associated with a larger imaginary part of the Floquet-Bloch wave vector, q.  
\par
We show in Fig. 10 the electric field profiles in the compact PC for the two aforementioned cases, versus the coordinate $x$, along the propagation direction; in (a) the case with $\textrm d_{\textrm{int}}=0.05 \hspace{0.01mm} \textrm d_1$ and in (b) the case with $\textrm d_{\textrm{int}}=0.5 \hspace{0.01mm} \textrm d_1$. We compare this profile with the Bloch-phase envelope for the electric amplitude $|\textrm E|$, $e^{-{\textrm{Im(q)}}x}$, shown as dotted lines in both sub-figures. The Bloch-envelope captures the relative phase for PC points that are spaced integer multiples of the lattice constant, a \cite{longprb, complexbands}. In other words, it provides a prediction for the electric field amplitude decay as the EM wave propagates from unit cell to unit cell of a semi-infinite PC. It is impressive to observe, how well this prediction captures the electric field decay from the front to the back layer for the case of Fig. 10(a). This is because the PC although compact, is reflectionless, so it emulates the propagation characteristics of its semi-infinite counterpart. This is not true however for the case of Fig. 10(b), hence the disagreement between Bloch-phase prediction and observed electric field amplitude decay.
\par 
In order to evaluate the performance of the proposed compact SiC PC-based absorber we compare it with that of a single SiC block in the same frequency regime. We show with dot-dashed line in Fig. 9 the absorption from bulk SiC, being more than a wavelength thick . Conversely, we show with dashed lines the absorption from an ultra-thin 12.5 nm-thick SiC film, which is as thick as the top layer of the structure of Fig. 9(d). The two extremes capture the bounds for the absorptance behavior of a SiC slab of intermediate thicknesses. For the ultra-thin slab, we do observe a peak of $\sim 40\%$ at 12.6 $\upmu$m free space wavelength, at the edge of the SiC Reststrahlen band. We emphasize that the physics of such an absorption peak for the single thin film slab is entirely different from the ones observed in the design of Fig. 9(a). In the latter cases the peaks are a photonic crystal effect while in the former it is a combination of thin-film behavior and the extreme SiC optical parameters at the Reststrahlen band-edge. This can in fact be easily checked by applying the thin-film approximation into the expressions \cite{voula2} for transmission, T, and reflection, R of an absorbing block of thickness $L$, and  complex refractive index $\textrm n+\textrm i \kappa$, where n and $\kappa$ are much larger than 1. 
\par
\par
\begin{figure}[!htbp] 
\begin{center}  
\includegraphics[angle=0, width=8.0cm]{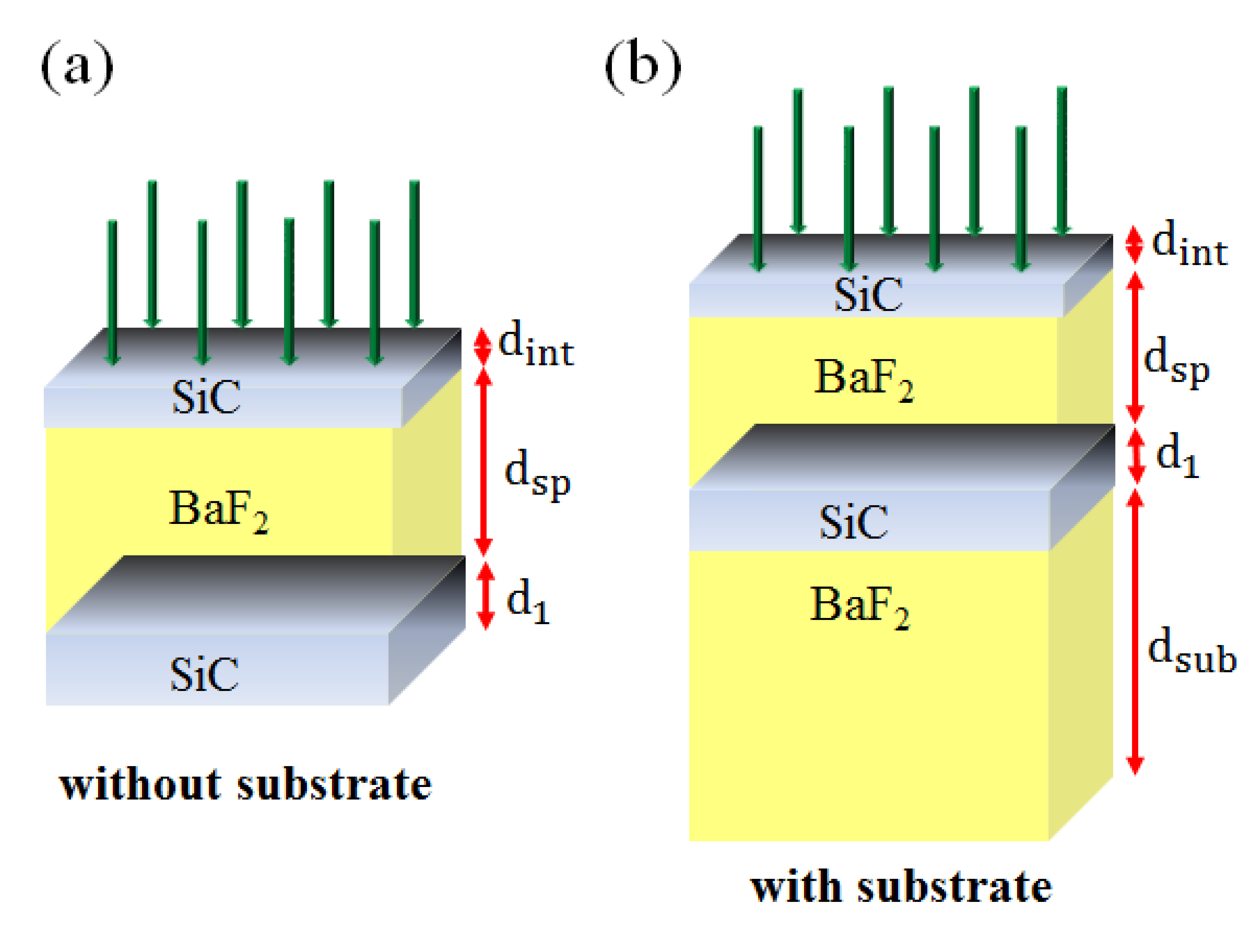}
\caption{(Color online) (a) Schematics of the realizable compact PC with all structural information indicated. (b) Same as the design in (a) but resting on a substrate made from the spacer material. }
\end{center}    
\end{figure}
\par   

We observe a very large absorption enhancement when compared to the capabilities of a single SiC slab, in both cases. It is interesting to check where is the power getting absorbed in our compact system. The filled circles in Fig. 10 depict the ratio of incident power that is absorbed in in each SiC layer of our proposed compact design. In both the cases of Fig. 10(a) and (b) we find that the thinner front slab absorbs more light. However, the case of Fig. 10(a) attest a truly astonishing absorbing phenomenon, where more than $90\%$ of the incident light gets absorbed by the front 12.5 nm-thick layer. This most extraordinary absorbing behavior would be highly attractive for applications. Thus we explore in the following section practically realizable designs.
\par
\section{Practically realizable superabsorber designs} 
\par
Here we investigate some variations of the compact superabsorber of Sec. IV, that would suggest potential for practical realization. We introduce a transparent material spacer instead of air, as can be seen in the schematics of Fig. 11(a). As an example for the transparent material spacer we consider $\textrm{BaF}_2$, which has a refractive index of $\sim 1.36$ in the frequency regime of interest \cite{BaF2}. In addition, we will investigate the influence of a substrate that is made from the same transparent material as the dielectric spacer [schematics of Fig. 11(b)]. In Figs. 11(a) and 11(b) we have identified symbolically all the pertinent geometric features, that we intend to ``tweak'' in the following with the aim towards superabsorbing behavior.  
\par 
\par
\par
\begin{figure}[!htbp] 
\begin{center}  
\includegraphics[angle=0, width=6.0cm]{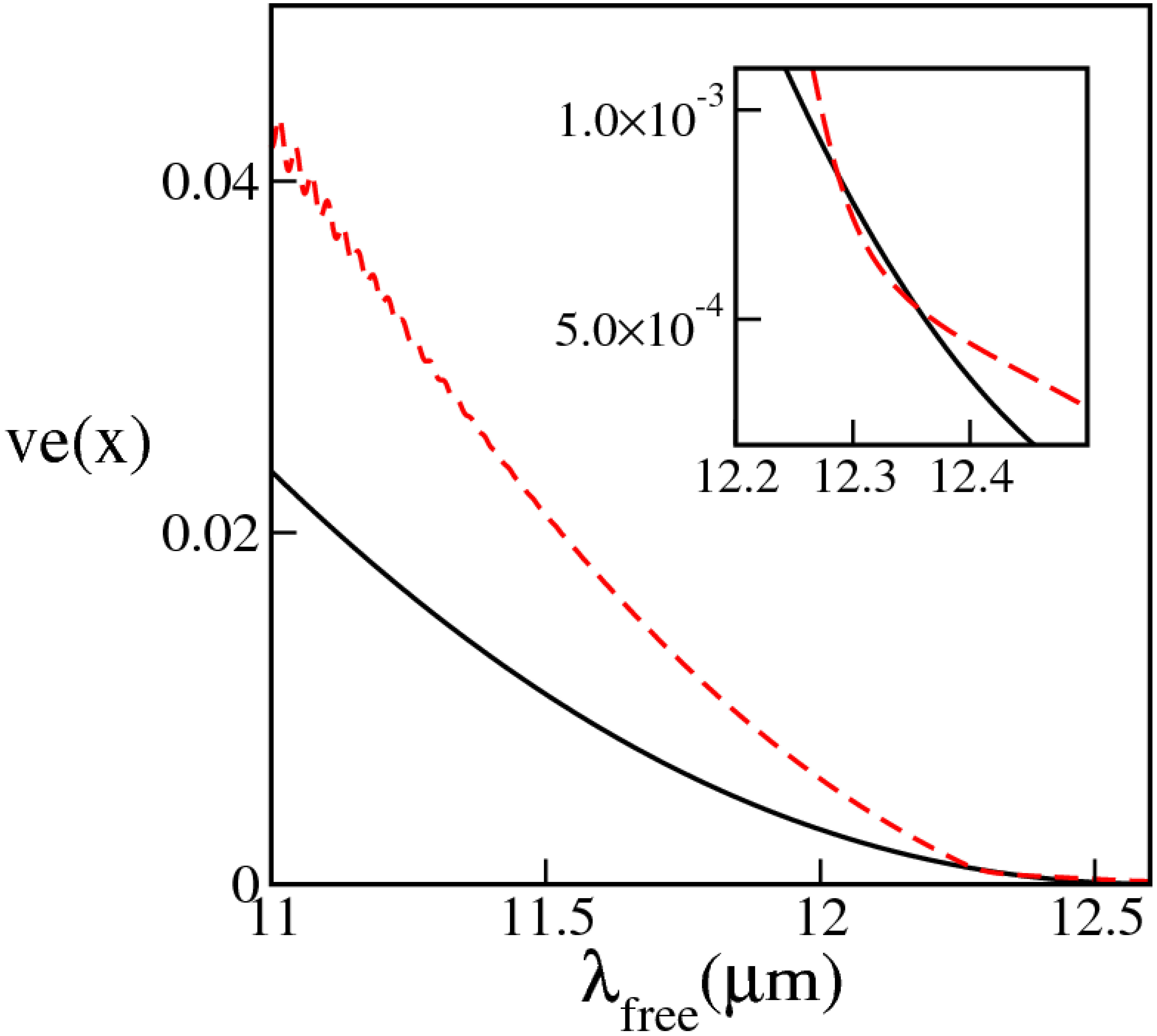}
\caption{(Color online) Energy velocity versus free space wavelength at the interface of a semi-infinite SiC-$\textrm{BaF}_2$ PC of lattice constant a=3.5 $\upmu$m and SiC filling ratio equal to 0.065 (dashed lines). The required optimum of Eq. (1) is shown with a solid line. The inset highlights the wavelength region where the interface energy velocity intersects with the required optimum value. }
\end{center}    
\end{figure}
\par 
In order to determine the parameters for which the designs of Fig. 11 can act as a superabsorber we go back again to the corresponding semi-infinite PC. i.e. we consider a SiC-$\textrm{BaF}_{2}$ photonic crystal and look into the spectral behavior of energy velocity at the interface,  ${\textrm v}_{\textrm{e,int}}$. We search for parameters that yield an intersection between ${\textrm v}_{\textrm{e,int}}$ at the theoretically mandated optimum ${\textrm v}_{\textrm e,0}$. Again, this intersection wavelength should be in the proximity of the PC band-edge, when targeting absorptance performance with the compact structure. 
\par 
\par
\begin{figure}[!htbp]    
\begin{center}  
\includegraphics[angle=0, width=8.0cm]{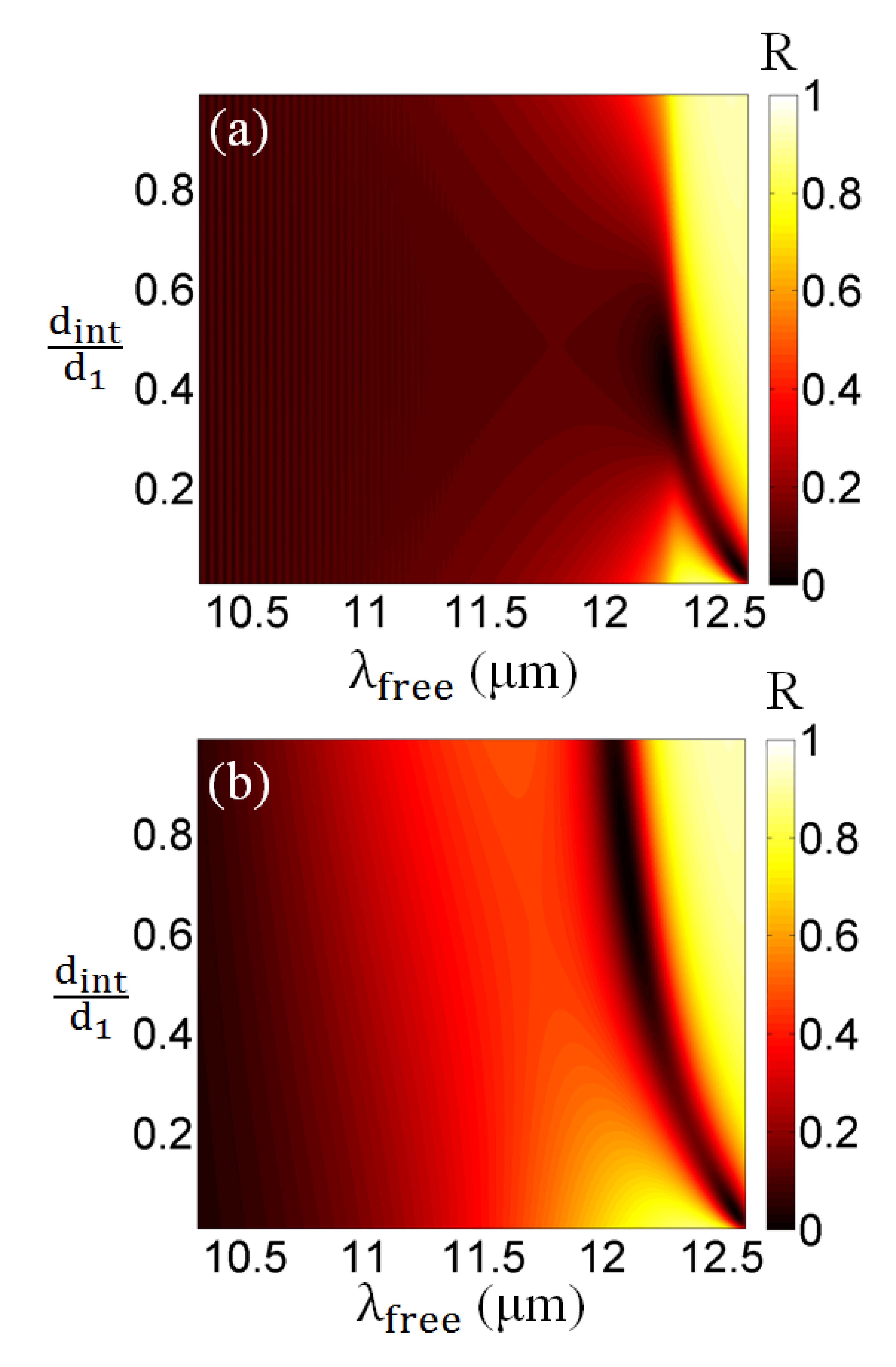}
\caption{(Color online) Reflectance (color-map), versus free space wavelength $\lambda_{\textrm{free}}$, and front-layer truncation ratio, $d_{\textrm{int}}$/$\textrm d_1$ for the SiC-$\textrm{BaF}_{2}$ system. In (a) the result of the semi-infinite PC is shown. In (b) the corresponding compact system of Fig. 11(a) is shown.}
\end{center}    
\end{figure} 
\par 
As a paradigm we present the case of filling ratio f=0.065 and lattice constant a=3.5 $\upmu$m, for which we plot the interface energy-velocity versus free space wavelength with dashed lines in Fig. 12. The required optimum of Eq. (1), ${\textrm v}_{\textrm e,0}$, is also indicated with a solid line. We can clearly identify two intersections between the ${\textrm v}_{\textrm{e,int}}$-${\textrm v}_{\textrm{0}}$ curves that are close to each other spectrally, (see blow-up of this frequency region in the inset), and in the proximity of the PC band-edge. This means that there are potentially two near-band-edge reflectionless frequencies. We note, that usually one can always find a termination for which both the energy velocity and the spatial gradient of energy velocity at the interface meet the mandated optimum values of Eqs. (1) and (2) at frequencies that nearly coincide, but this may not always happen. One should also keep in mind that the ${\textrm v}_{\textrm{e,int}}$ values may shift somewhat with termination as well. This effect is offcourse stronger for the smaller termination ratios. 
\par     
In this SiC-$\textrm{BaF}_2$ PC we find two regimes of near-reflectionless behavior;  one near a half-terminated front-layer and  another for a  very large front-layer truncation. These can be seen as the very dark regions in the reflectance map of Fig. 13(a), where reflectance, R, is plotted versus free space wavelength, $\lambda_{\textrm{free}}$, and front-layer termination ratio $\textrm d_{\textrm{int}}$/$\textrm d_1$. For comparison, the reflectance for the compact structure of Fig. 11 is also shown in Fig. 13(b). Indeed, these near-reflectionless $\lambda_{\textrm{free}}$-$\textrm d_{\textrm{int}}$/$\textrm d_1$regimes are quite robust with shrinking PC size . In particular, we observe that both near-reflectionless regimes survive until the extreme case of the compact three-layer structure, where the near-zero-reflection property emerges at front-layer truncation ratios of 0.05 and 0.50. 
\par 
We show for these above mentioned cases the absorptance, A and reflectance, R in Figs. 14(a) and 14(b) respectively. The solid lines represent the case of a 0.05 truncation ratio, while the dashed lines represent the case of a 0.50 truncation ratio. The corresponding symbols, -circles and diamonds-, represent the respective result when the compact three-layer structure is placed on a forty-micron-thick substrate. We observe that the substrate has almost no influence at all to the absorptance. This is a promising result towards the realization of the compact absorber system depicted in Fig. 11.
\par
\begin{figure}[!htbp] 
\begin{center}    
\includegraphics[angle=0, width=7.0cm]{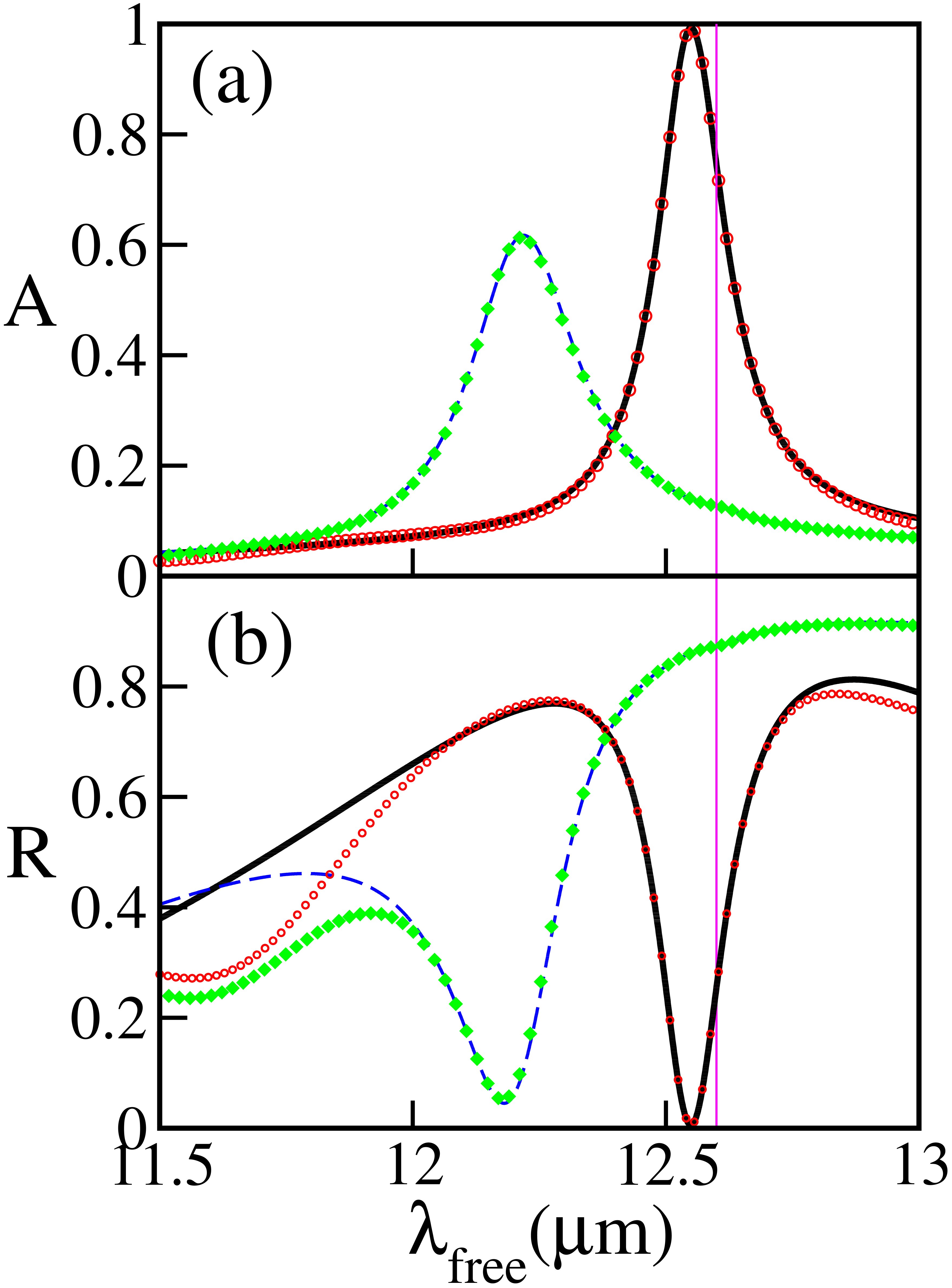} 
\caption{(Color online) Absorptance [(a)] and Reflectance [(b)] versus free space wavelength $\lambda_{\textrm{free}}$ for the compact SiC-$\textrm{BaF}_2$-SiC system corresponding to a PC with a lattice constant a=3.5 $\upmu$m and a SiC filling ratio of f=0.065. Two cases of front-layer truncation are shown: the case of 0.05 truncation ratio with solid lines and the case of 0.5 truncation ratio with dashed lines. The corresponding circles and diamonds represent the respective result when the compact three-layer system rests on a 40$\upmu$m thick substrate made of $\textrm{BaF}_2$. }
\end{center}     
\end{figure}  
\par  
\begin{table}
\caption{\label{tab:table2} Outline of performance of the compact superabsorber of Fig. 11(a) for two front-layer truncation ratios [A stands for absorptance, and DPR stands for dissipated power ratio]}
\begin{ruledtabular}
\begin{tabular}{cccccccc}
 $Truncation$ &$Total\ A$ &$DPR$&
 $DPR$  &$ \%\ of\ total\ A$ &$\%\ of\ total\ A$ \\
 $Ratio$ &$at\ peak\ wavelength$ &$in\ front\ layer$& $in\ back\ layer$
 &$in\  front\ layer$  &$ in\  back\ layer$ &$$\\
\hline
$\textrm d_{\textrm{int}}/\textrm d_1=0.05$& 0.993 & 0.925 & 0.0680 & 0.93 & 0.068  \\
$\textrm d_{\textrm{int}}/\textrm d_1=0.5$& 0.617 & 0.428 & 0.19 & 0.69 & 0.30
\end{tabular}
\end{ruledtabular} 
\end{table}
We find a remarkable absorption performance for the compact design for both truncation ratios, which we describe in more detail in Table I. We were able to observe also with the  realizable SiC-$\textrm{BaF}_{2}$ the extra-ordinary near-perfect absorption, that we saw in previous section in the SiC-air system. Our proposed compact SiC-$\textrm{BaF}_{2}$ system has a total thickness of about $\sim \lambda_{\textrm{free}}/3$. At the wavelength of near-perfect absorption, more than $92\%$ of the incoming light is getting absorbed by the top layer, which is less than $\sim \lambda_{\textrm{free}}/1000$ thick. This is a truly astonishing performance for this simple three-layer paradigm. It should be noted, that we found that by tweaking the geometric parameters of our proposed compact design for a reflectionless condition slightly away from the band-edge, it is possible to obtain an extra-ordinary absorption enhancement that depending on front-layer truncation can swap all the phonon-polariton band-gap region; however near-perfect absorption was not reached in such a case for any front-layer truncation.   
\par 
To recap, we demonstrated here a realizable compact superabsorber where almost all the incoming light is absorbed by the top layer which is less than a thousand times the wavelength thick. The superabsorber comprises of a highly absorbing (large $\kappa$) material and a transparent medium. We note, in passing that very recently compact absorbers in a planar geometry have been also reported by M. A. Kats et al. \cite{capasso} and W. Streyer et al. \cite{streyer}, with designs that rely on the mutual optical properties of two lossy materials in the former work or a high index dielectric and an engineered metal of $\varepsilon \sim -1$ in the latter work. In our proposed compact design the underlying mechanism is a PC effect, where absorption is facilitated by a reflectionless funneling to a photonic-crystal mode, with a highly lossy Floquet-Bloch phase. This mechanism leads to compact structures made of a single-kind highly lossy medium in a planar geometry, and any transparent medium, that can function as highly efficient absorbers by suitable adjustment of the out-of-plane constituent slab thicknesses.  
\section{Conclusion}  
We have presented here a new paradigm of a practically realizable SiC-$\textrm{BaF}_{2}$ layered system. We demonstrated with a compact design, which is a third-of-the wavelength thick, a near-perfect absorption, where more than $92 \%$ of the incoming light is absorbed in the top SiC ultra-thin layer, having a thickness a thousand times less than the impinging wavelength. The reported super-absorber effect emanates from a special photonic-crystal behavior of the corresponding photonic lattice. In particular, the underpinning mechanism is the achievement of near-zero reflection in the proximity of the photonic 1D lattice band-edge, by a special truncation of the front layer. This enables funneling all incoming light into an ultra-slow mode, that allows the rapid decay of the EM fields within the PC, manifested by the large imaginary part of the complex Floquet-Bloch phase. We believe, our proposed platform offers new avenues for absorption harnessing across the EM spectrum and will inspire new designs where absorption occurs in a one-step process without anti-reflection coating and/or back-reflector \cite{fan} and within a single kind of highly absorbing material.  
\section{Acknowledgments}    
\par
Financial support for the Ph.D. studentship of G. C. R. Devarapu by the College of Engineering, Mathematics and Physical Sciences (CEMPS) University of Exeter is acknowledged.

\newpage
\begin{center}

{\bf \large Captions}
\vspace{2mm}
\end{center}
\par
{\bf Fig. 1}{(Color online) (a) Schematics of the SiC-air 1D-PC with the geometric parameters indicated. (b) Spectral response of the real (solid) and imaginary (dashed) parts of  the SiC permittivity model of Eq. (3). (c) Absorption (solid line) and reflection (dashed line) for a thick bulk SiC block.}
\par
{\bf Fig. 2}{(Color online) Spectral response of the energy velocity at the interface ${\textrm v}_{\textrm{e,int}}$ of a semi-infinite SiC-air PCs structure is shown as solid lines. The dashed lines depict the corresponding values for the same PCs but with $50\%$ of their entry face being cut-off. The results in (a), (b) and (c) represent the PC cases with a lattice constant of a equal to 5 $\upmu {\textrm m}$, 8 $\upmu {\textrm m}$ and 10 $\upmu {\textrm m}$, respectively. In all cases, the interface-energy velocity value of the reflectionless condition, ${\textrm v}_{\textrm{e,0}}$ of Eq. (1), is depicted with dotted lines. Note, all energy velocity values are expressed in terms of the speed of light $c$. The vertical solid lines represent the spectral position of the absorption peaks that we will observe in Fig. 5.}
\par
{\bf Fig. 3}{(Color online) The energy-velocity gradient is shown for two PC systems with a lattice constant equal to 5 $\upmu {\textrm m}$ and 10 $\upmu {\textrm m}$ in panels (a) and (b) respectively. The horizontal dashed line represents the reflectionless condition value dictated by Eq. (2). Note the coordinate within the PC entry layer, $x-$, is expressed in terms of the lattice constant a,  while the energy velocity gradient is expressed in terms of $c/\textrm a$, with $c$ being the speed of light.} 
\par
{\bf Fig. 4}{(Color online) Reflection (in color-map) versus termination ratio, $\textrm d_{\textrm{int}}/\textrm d_1$ and free space wavelength, $\lambda_{\textrm {free}}$ calculated from TMM. Panels (a) and (b) represent the result corresponding to the semi-infinite PCs with lattice constant a, of 5  $\upmu {\textrm m}$, and 10 $\upmu {\textrm m}$ respectively. Same is shown in (c) and (d) but for 200 $\upmu$m-thick PCs.}
\par
{\bf Fig. 5}{(Color online) Absorptance versus free space wavelength, $\lambda_{\textrm{free}}$, for three 200 $\upmu {\textrm m}$ thick SiC-air PCs of 0.05 filling ratio and $50\%$ front layer truncation. The solid, dashed and dot-dashed curves correspond to PCs with a lattice constant a equal to 10 $\upmu {\textrm m}$, 8 $ \upmu {\textrm m}$, and 5 $\upmu {\textrm m}$ respectively. The front SiC layer is terminated to half its original size.} 
\par
{\bf Fig. 6}{(Color online) Complex band structure (free space wavelength versus Bloch wavevector q) for the PC cases of lattice constant a, 5 $\upmu {\textrm m}$ [in (a) and (b)] and 10 $\upmu {\textrm m}$ [in (c) and (d)] . The respective reflectionless-condition wavelengths are indicated with horizontal dashed lines. Note, both the real and imaginary parts of the Bloch wave vector q is expressed in terms of $\pi/\textrm a$.} 
\par
{\bf Fig. 7}{(Color online) Dissipated to incident power ratio versus free space wavelength, $\lambda_{\textrm{free}}$, for the 200$\upmu$m thick SiC-air PCs with $50\%$ truncated front layer, within the first $\textrm N_c$ PC unit cells. The result in (a) [(b)] corresponds to the PC case with 5 $\upmu {\textrm m}$ [10 $\upmu {\textrm m}$] lattice constant. The respective absorptance is shown for reference with the dark solid line. Note, the total number of PC unit cells, N, is 40 for the case in (a) and 20 for the case in (b).} 
\par
{\bf Fig. 8}{(Color online) Absorptance enhancement of the two terminated SiC-air PCs with  lattice constant a, 5 $\upmu {\textrm m}$ (dot-dashed line with diamonds),  and 10 $\upmu {\textrm m}$ (solid line with filled circles)  with respect to the absorption of a SiC block about a wavelength-thick is plotted against the total thickness of SiC  encountered by the EM wave as it travels through the PC.}
\par
{\bf Fig. 9}{(Color online) (a) Schematics of the compact PC-based design with all structural information indicated. (b) Reflectance (color-map) versus free space wavelength $\lambda_{{\textrm{free}}}$ and front-layer truncation ratio $\textrm d_{{\textrm{int}}}/{\textrm d_1}$. (c) Absorptance (solid lines) and reflectance (dotted lines), for the design in (a) with (c) [(d)] showing the case of $\textrm d_{\textrm{int}}/{\textrm d_1}=0.5$ [$\textrm d_{\textrm{int}}/{\textrm d_1}=0.05$]. For comparison absorptance through a single layer is also shown for bulk SiC (dot-dashed) and an ultra-thin SiC film as thick as the front layer of the structure of Fig. 9(d). The vertical line designates the SiC Reststrahlen band-edge.}
\par
{\bf Fig. 10}{(Color online) Electric field amplitude, $|\textrm E|$, profiles (left vertical axis) versus the coordinate $x$ within the compact superabsorber design. The depicted profiles are normalized with the incident electric field amplitude $|\textrm E_0|$. The dotted lines represent the $|\textrm E|$-decay, from the front to theback later, as predicted by the complex band structure of Fig. 6. The solid circles represent the ratio of incident power that is absorbed in each layer (see right vertical axis for values). Panel (a) and (b) represent the respective cases with front-to-back-layer truncation ratio of 0.05 and 0.5.}
\par
{\bf Fig. 11}{(Color online) (a) Schematics of the realizable compact PC with all structural information indicated. (b) Same as the design in (a) but resting on a substrate made from the spacer material. }
\par
{\bf Fig. 12}{(Color online) Reflectance (color-map), versus free space wavelength $\lambda_{\textrm{free}}$, and front-layer truncation ratio, $d_{\textrm{int}}$/$\textrm d_1$ for the SiC-$\textrm{BaF}_{2}$ system. In (a) the result of the semi-infinite PC is shown. In (b) the corresponding compact system of Fig. 11(a) is shown.}
\par
{\bf Fig. 13}{(Color online) Reflectance (color-map), versus free space wavelength $\lambda_{\textrm{free}}$, and front-layer truncation ratio, $d_{\textrm{int}}$/$\textrm d_1$ for the SiC-$\textrm{BaF}_{2}$ system. In (a) the result of the semi-infinite PC is shown. In (b) the corresponding compact system of Fig. 11(a) is shown.}
\par
{\bf Fig. 14}{(Color online) Absorptance [(a)] and Reflectance [(b)] versus free space wavelength $\lambda_{\textrm{free}}$ for the compact SiC-$\textrm{BaF}_2$-SiC system corresponding to a PC with a lattice constant a=3.5 $\upmu$m and a SiC filling ratio of f=0.065. Two cases of front-layer truncation are shown: the case of 0.05 truncation ratio with solid lines and the case of 0.5 truncation ratio with dashed lines. The corresponding circles and diamonds represent the respective result when the compact three-layer system rests on a 40$\upmu$m thick substrate made of $\textrm{BaF}_2$. } 
\par

\begin{thebibliography}{599}
\bibitem{semi-infinite}
We emulate a semi-infinite PC behavior by considering a thick two-hundred-cell structure. In this way, essentially all light is absorbed when it meets the end-face, and backreflection, which is inevitable in finite systems, is eliminated. See also Ref. \onlinecite{longprb}
\end{thebibliography}
\end{document}